\documentclass[a4paper,fleqn,usenatbib]{mnras}

\usepackage{newtxtext,newtxmath}

\usepackage[T1]{fontenc}
\usepackage{ae,aecompl}

\usepackage{graphicx}	% Including figure files
\usepackage{amsmath}	% Advanced maths commands
\usepackage{amssymb}	% Extra maths symbols
\def\fluxunits{erg\,cm$^{-2}$\,s$^{-1}$}

\def\lumunits{erg\,s$^{-1}$}
\def\reducedchi{$\chi_{\mathrm{red}}^{2}/\mathrm{d.o.f} $}
\def\XMM{\textit{XMM-Newton}~}
\def\INTEGRAL{\textit{INTEGRAL}~}

\title[\XMM and \INTEGRAL analysis of IGR J17354-3255]{\XMM and \INTEGRAL analysis of the Supergiant Fast X-ray Transient IGR J17354-3255}

\author[M. E. Goossens et al.]{
M.~E.~Goossens,$^{1}$\thanks{E-mail: M.E.Goossens@soton.ac.uk}
A.~J.~Bird,$^{1}$
A.~B.~Hill,$^{1}$
V.~Sguera,$^{2}$
and {S.~P.~Drave}$^{1}$
\\
$^{1}$School of Physics and Astronomy, Faculty of Physical Sciences and Engineering, University of Southampton, SO17 1BJ, UK\\
$^{2}$INAF-OAS, Osservatorio di Astrofisica e Scienza dello Spazio di Bologna, Area della Ricerca del CNR, via Gobetti 101, 1-40129 Bologna, Italy\\
}

\date{Accepted XXX. Received YYY; in original form ZZZ}

\pubyear{2015}

\hypersetup{draft}
\begin{document}
\label{firstpage}
\pagerange{\pageref{firstpage}--\pageref{lastpage}}
\maketitle

\begin{abstract}
We present the results of combined \INTEGRAL and \XMM observations of the supergiant fast X-ray transient (SFXT) IGR J17354$-$3255. Three \XMM observations of lengths 33.4 ks, 32.5 ks and 21.9 ks were undertaken, the first an initial pointing to identify the correct source in the field of view and the latter two performed around periastron. Simultaneous \INTEGRAL observations across $\sim66\%$ of the orbital cycle were analysed but the source was neither detected by IBIS/ISGRI nor by JEM-X.
The \XMM light curves display a range of moderately bright X-ray activity but there are no particularly strong flares or outbursts in any of the three observations. We show that the spectral shape measured by \XMM can be fitted by a consistent model throughout the observation, suggesting that the observed flux variations are driven by obscuration from a wind of varying density rather than changes in accretion mode. The simultaneous \INTEGRAL data rule out simple extrapolation of the simple powerlaw model beyond the \XMM energy range.
\end{abstract}


%%%%%%%%%%%%%%%%%%%%%%%%%%%%%%%%%%%%%%%%%%%%%%%%%%

%%%%%%%%%%%%%%%%% BODY OF PAPER %%%%%%%%%%%%%%%%%%

\section{Introduction}

Supergiant Fast X-ray Transients (SFXTs) consist of a compact object, which is either a neutron star or black hole, orbiting a supergiant donor and accreting from its wind \citep{Sguera2005, Sguera2006, Negueruela2005}. These sources are a subclass of high-mass X-ray binaries (HMXBs) which are occasionally observed in X-ray quiescence ($L_{x} < 10^{32}$ erg$\textrm{s}^{-1}$) but more frequently during a low X-ray state. Occasionally, fast X-ray transient activity can be observed from these binary systems, this being characterised by outbursts lasting less than a day and flares of a few tens of minutes and peak luminosities of $\sim 10^{36}$ erg~$\textrm{s}^{-1}$ similar to those of persistent supergiant X-ray binaries (SGXBs) \citep{Walter2007}.

SFXT systems are generally identified by their low persistent luminosity and their high dynamic range on variable timescales ($\sim10^{3}-10^{5}$ times the range observed in classical systems (\citet{Sguera2005}, \citet{Romano2015}). The origin of this behaviour is still a matter of debate, with several viable models used to explain this phenomenon. These models range from accretion of dense inhomogeneous ``clumps'' in the winds of the supergiant companions (\citet{Ducci2009b}, \citet{In'tZand2005}), to the presence of magnetic/centrifugal gates generated by the neutron stars \citep{Bozzo2008b}, to a subsonic settling accretion regime combined with magnetic reconnections \citep{Shakura2014}.

IGR J17354$-$3255 was discovered in April 2006 as an unidentified hard X-ray transient during a monitoring observation of the Galactic bulge region \citep{Kuulkers, Kuulkers2007} with the INTErnational Gamma-Ray Astrophysics Laboratory ({\em INTEGRAL}). The source reached a flux of $\sim 2.1\times10^{-10}$\fluxunits{ }(18 mCrab) in the 20--60 keV band during two consecutive 1.8ks pointings and was located towards the Galactic Center. IGR J17354$-$3255 is reported in the latest IBIS catalogue \citep{Bird2016} as a persistent source with an average flux of $8.33\times10^{-12}$\fluxunits (1.1 mCrab) in the 20--40 keV energy band. 

IGR J17354$-$3255 is also listed in the \textit{Swift} Burst Alert Telescope (BAT) \citep{Barthelmy2005} 58-month Hard X-ray Survey \citep{Baumgartner2010} with a 14--195 keV average flux of $2.7\times10^{-11}$\fluxunits as well as the 54 month Palermo \textit{Swift}/BAT hard X-ray catalogue \citep{Cusumano2010} with a 15--150 keV average flux of $2.1\times10^{-11}$\fluxunits.

Due to its hard X-ray flaring activity with a mean flaring flux of $2.6-5.2\times10^{-12}$\fluxunits{ }(20--40 mCrab) and peak flaring flux of $1.4\times10^{-9}$\fluxunits{ }(108 mCrab) in the 18--60 keV energy band, together with its large dynamic range, \citet{Sguera2011} proposed IGR J17354$-$3255 as a candidate intermediate supergiant fast X-ray transient (SFXT) much like IGR J16465$-$4507 \citep{Clark2010a} and several others \citep{Walter2007, Sguera2007}. The X-ray dynamic range was further increased by a non-detection during observations with \XMM \citep{Bozzo2012b}, giving  support for the classification of this source as an SFXT.

\citet{DAi2011} and \citet{Sguera2011} studied the hard X-ray properties of IGR J17354$-$3255 using archival \textit{Swift}/BAT and \INTEGRAL observations respectively. They discovered a periodic signal of $8.447\pm0.002$ d, which they interpreted as the orbital period of the binary system.

\textit{Chandra} observations \citep{Tomsick2009c} reduced the X-ray error box of the source to arcsec size consequently allowing identification of the optical/infrared counterpart. \citet{Coleiro2013a} subsequently measured the nIR $K_{\mathrm{s}}$-band spectrum for IGR J17354$-$3255 suggesting that the supergiant spectral type is likely an O8.5Iab(f) or an O9Iab, further confirming the sgHMXB/SFXT classification. Its high optical/IR extinction and the position within 5 degrees of the Galactic Center may indicate a distance near 8.5 kpc \citep{Tomsick2009atel}.

IGR J17354$-$3255 has been  proposed as the best candidate counterpart of the gamma-ray transient AGL J1734$-$3310 based on spatial correlation as well as on a similar flaring nature on short timescales \citep{Sguera2011, Sguera2013}. IGR~J17354$-$3255 is the only hard X-ray source that can  be  unambiguously  said  to  be  located  within  the 0.46 degrees error circle radius of AGL J1734$-$3310. The latter is an 
MeV transient source discovered by \textit{AGILE}/GRID in 2009 April during a 1 day duration flare \citep{Bulgarelli2009}.

More recently, \cite{Ducci2013} performed the first full-orbital monitoring of this source with \textit{Swift}/XRT and \cite{Bozzo2017}  included IGR J17354$-$3255 in a comprehensive \XMM study of several SFXTs aiming to investigate their accretion environment.

In this paper, we report on three \XMM observations together with a simultaneous \INTEGRAL observation of IGR J17354$-$3255 where the source exhibits typical SFXT behaviour with a high level of flaring activity, although no particularly bright outbursts are seen. 

\section{Observations and Data Reduction}

\subsection{\INTEGRAL}

The \INTEGRAL gamma-ray observatory consists of three co-aligned coded mask telescopes: the soft X-ray monitor (JEM-X) \citep{Lund2003}, the spectrometer (SPI) \citep{Vedrenne2003} and the hard X-ray imager (IBIS) \citep{Ubertini2003}. \INTEGRAL observations are usually divided into science windows that have a duration of $\sim2000$s. Data from IBIS and JEM-X were analysed using the `Off-line Scientific Analysis' (OSA) software version 10.0.

\begin{figure}
	\begin{center}
		\includegraphics[width=\columnwidth]{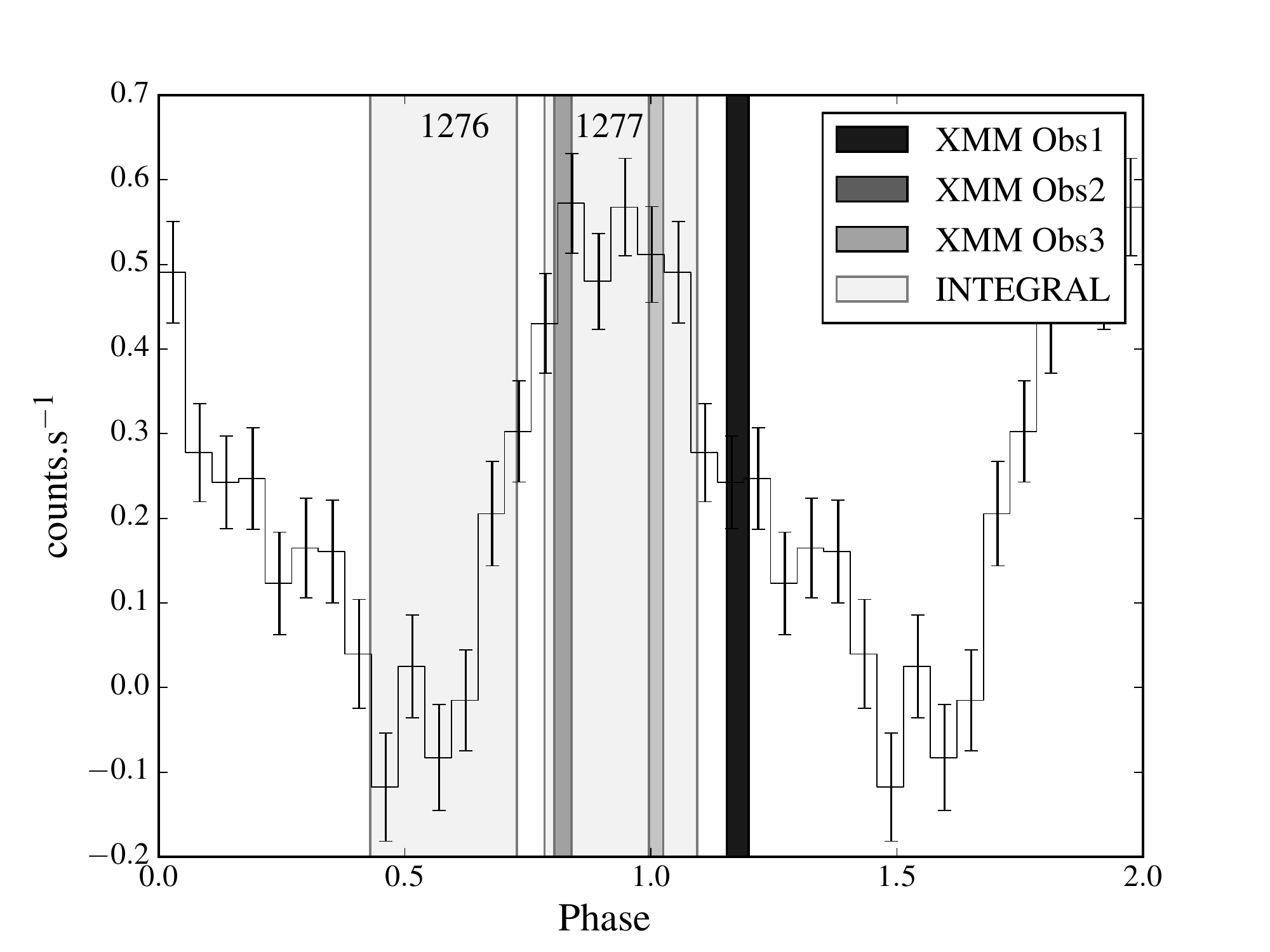}
		\caption{\label{fig:phasefold} Phase folded light curve of the IBIS/ISGRI data set for IGR J17354$-$3255 in the 18-60 keV energy band with the ephemeris MJD 55925.895 as phase=0 and a period of 8.447 days. The durations of the three \XMM observations as well as the 327 ks on-source \INTEGRAL coverage are indicated according to phase. The two \INTEGRAL revolutions are marked by their respective numbers.}
	\end{center}
\end{figure}

IGR J17354$-$3255 was in the IBIS/ISGRI fully coded field of view (FOV) (during a hexagonal dither pointing) for a total exposure of $\sim145$ ks during a proprietary targeted observation (PI Drave), which formed part of revolution 1277. However, by chance it was also in the FOV during the remaining part of revolution 1277 (for an additional total on-source exposure of $\sim$54 ks) as well as during the previous revolution 1276 (for an on-source exposure of $\sim130$ ks). Combining all the data from the two revolutions results in a total on-source exposure of 329 ks meaning that about 66$\%$ of an entire orbital cycle of IGR J17354$-$3255 is covered (Figure \ref{fig:phasefold}). 

\subsection{\XMM}

\begin{table*}
\caption{\XMM observation log for IGR J17354-3255. The effective exposure corresponds to the exposure assigned to the source position after filtering and cleaning processes.}
\label{table:one}
\centering
\begin{minipage}{170mm}
\begin{tabular}{c|c|c|c|c|c|c|ccc}
\hline
\hline
\multicolumn{1}{|c|}{Obs. No.}&\multicolumn{1}{|c|}{Date}&\multicolumn{1}{|c|}{S/C Rev.}&\multicolumn{1}{|c|}{Obs. ID}&\multicolumn{1}{|c|}{Start (UT)}&\multicolumn{1}{|c|}{Stop (UT)}&\multicolumn{1}{|c|}{Tot. Exp. (ks)}&\multicolumn{3}{|c|}{Eff. Exp. (ks)}\\
\hline
&&&&&&&EPIC-MOS1&EPIC-MOS2&EPIC-pn\\
1&15 Mar. 2013&2429&0693900201&13:04:56&22:21:52&33.416&33.076&33.128&0\\
2&29 Mar. 2013&2436&0701230101&11:48:26&20:50:44&32.538&25.101&25.116&24.082\\
3&31 Mar. 2013&2437&0701230701&03:35:07&08:40:24&21.917&21.630&21.635&19.685\\
\hline
\end{tabular}
\end{minipage}
\end{table*}

\begin{table*}
\caption{Source and background regions. In each case, the co-ordinates of the center of the region are given and the radius (or range of radii for annular regions) is specified in physical pixels.}
\label{table:sourcebackground}
\centering
\begin{minipage}{170mm}
\resizebox{\columnwidth}{!}{
\begin{tabular}{c|ccc|ccc}
\hline
\hline
\multicolumn{1}{|c|}{Observation}&\multicolumn{3}{|c|}{Source Region}&\multicolumn{3}{|c|}{Background Region}\\
\hline
&EPIC-MOS1&EPIC-MOS2&EPIC-pn&EPIC-MOS1&EPIC-MOS2&EPIC-pn\\
1&(22675.2, 22446.4, 200$-$1340)&(22678.3, 22459, 200$-$1420)&-&(28060.5, 28660.5, 2300)&(27980.5, 28620.5, 2100)&-\\
2&(24928, 24116.5, 1020)&(24945.9, 24126.5, 1100)&(24943.8, 24112, 400)&(22310.5, 26470.5, 1000)&(22490.5, 26170.5, 1000)&(22680.5, 29640.5, 1000)\\
3&(24965.3, 24104.4, 1420)&(24970.5, 24103.1, 1200)&(24976.8, 24093.8, 400)&(22330.5, 26250.5, 1000)&(24340.5, 26220.5, 800)&(22780.5, 29860.5, 1000)\\
\end{tabular}}
\end{minipage}
\end{table*}

IGR J17354$-$3255 was observed with \XMM in three observations between 2013 March 15 and March 31; details are given in Table \ref{table:one}. The \textit{XMM-Newton Observatory} consists of three $1500\textrm{cm}^{2}$ X-ray telescopes each carrying an European Photon Imaging Camera (EPIC) at the focus. Two of the EPICs use Metal Oxide Semi-conductor (MOS) CCDs \citep{Turner2001} and one uses a pn-CCD \citep{Strder2001a}. Additionally, there are Reflection Grating Spectrometer (RGS) \citep{DenHerder2001} arrays located behind two of the telescopes. 

The first observation was performed as an initial pointing to identify the correct source in the field of view (FOV). As a result IGR J17354$-$3255 was only observed with the MOS1 and MOS2 cameras but not with the PN. This was corrected for the consecutive observations so that the source was in the centre of the FOV. To schedule these observations an ephemeris of MJD 55924.959 was used as this was the best estimate of the periastron location at the time. We have subsequently calculated a refined ephemeris MJD 55925.895 using the 12 years of \INTEGRAL data available for this source. This first observation was thus performed at an orbital phase range of $\phi=0.154-0.199$. Observations two and three were performed at phase ranges $\phi=0.804-0.839$ and $\phi=0.996-0.025$, respectively, so shortly before and approximately at periastron, using this refined ephemeris as phase=0 and a period of 8.447 days (Figure \ref{fig:phasefold}). In observation one both EPIC-MOS cameras were operated in full-frame mode and in observations two and three both EPIC-MOS cameras as well as the EPIC-pn camera were operated in large window mode. In each case medium thickness filters were used in case of intense flaring behaviour. 

Observation data files (ODFs) were processed to produce calibrated event lists using the Science Analysis System (SAS) v15.0.0. RGS data was processed using the {\scshape rgsproc} task but the resulting FITS files had no counts for IGR J17354$-$3255 so that no further analysis was carried out for these instruments. We reprocessed the EPIC-MOS and EPIC-pn  events files with the {\scshape emproc} and {\scshape epproc} tasks, respectively. The data sets were checked for regions of high particle background following the method outlined in the \XMM SAS data analysis threads \citep{Gabriel2004}\footnote{https://www.cosmos.esa.int/web/xmm-newton/sas-threads}. We created single-event ({\scshape pattern == 0}) light curves above 10 keV for EPIC-MOS1 and EPIC-MOS2 and between 10 and 12 keV for pn covering the full FOV of each instrument using cut-offs between 0.2 and 0.4 counts $\textrm{s}^{-1}$ as appropriate. The good time intervals (GTIs) that remained after this filtering process were used in {\scshape evselect} to produce filtered event files, resulting in the relevant effective exposure times reported in Table \ref{table:one}. 

\begin{figure}
\minipage{0.4\textwidth}
	\centering
	\includegraphics[width=\columnwidth]{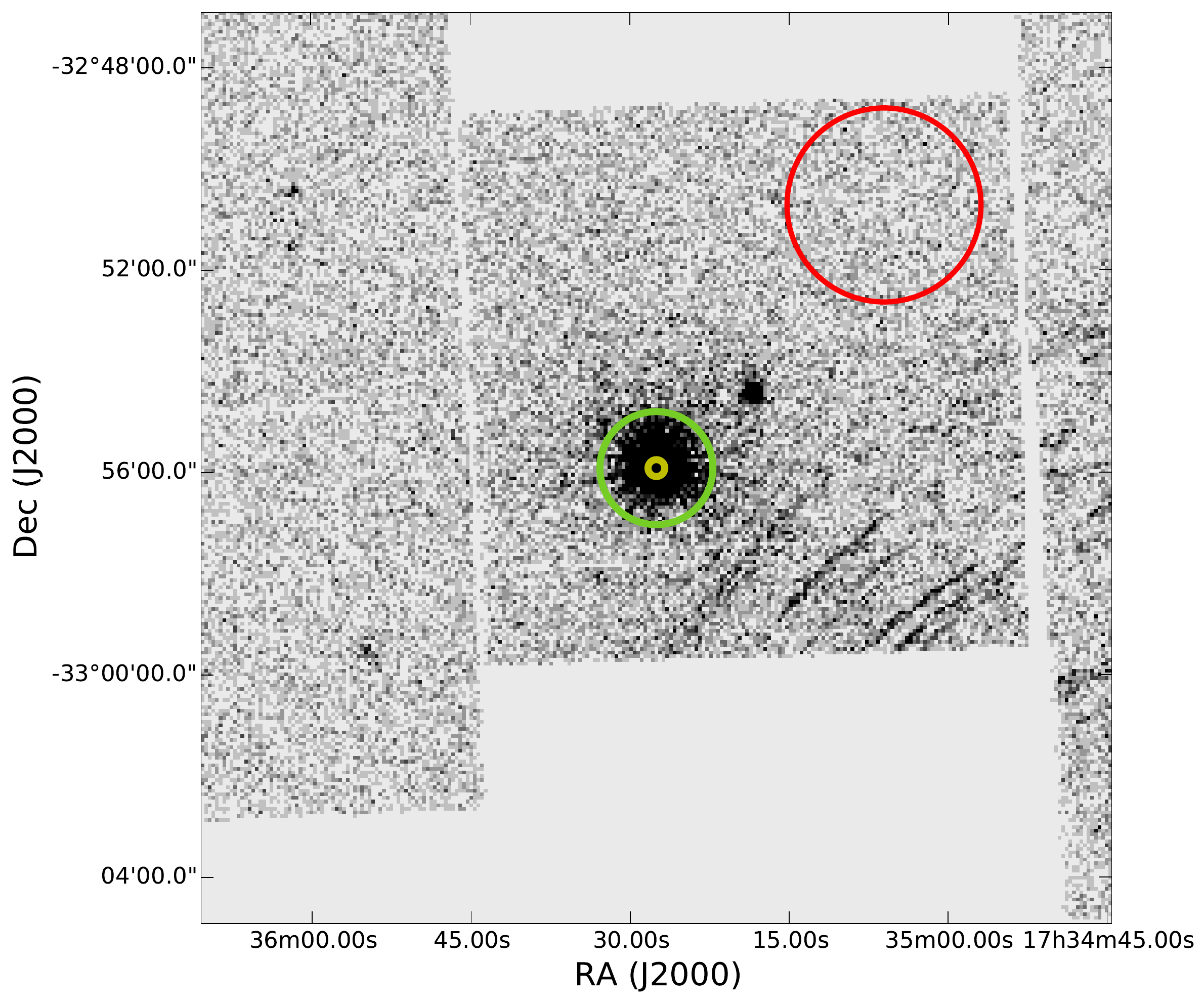}
	
\endminipage\hfill\newline
\minipage{0.4\textwidth}
	\centering
	\includegraphics[width=\columnwidth]{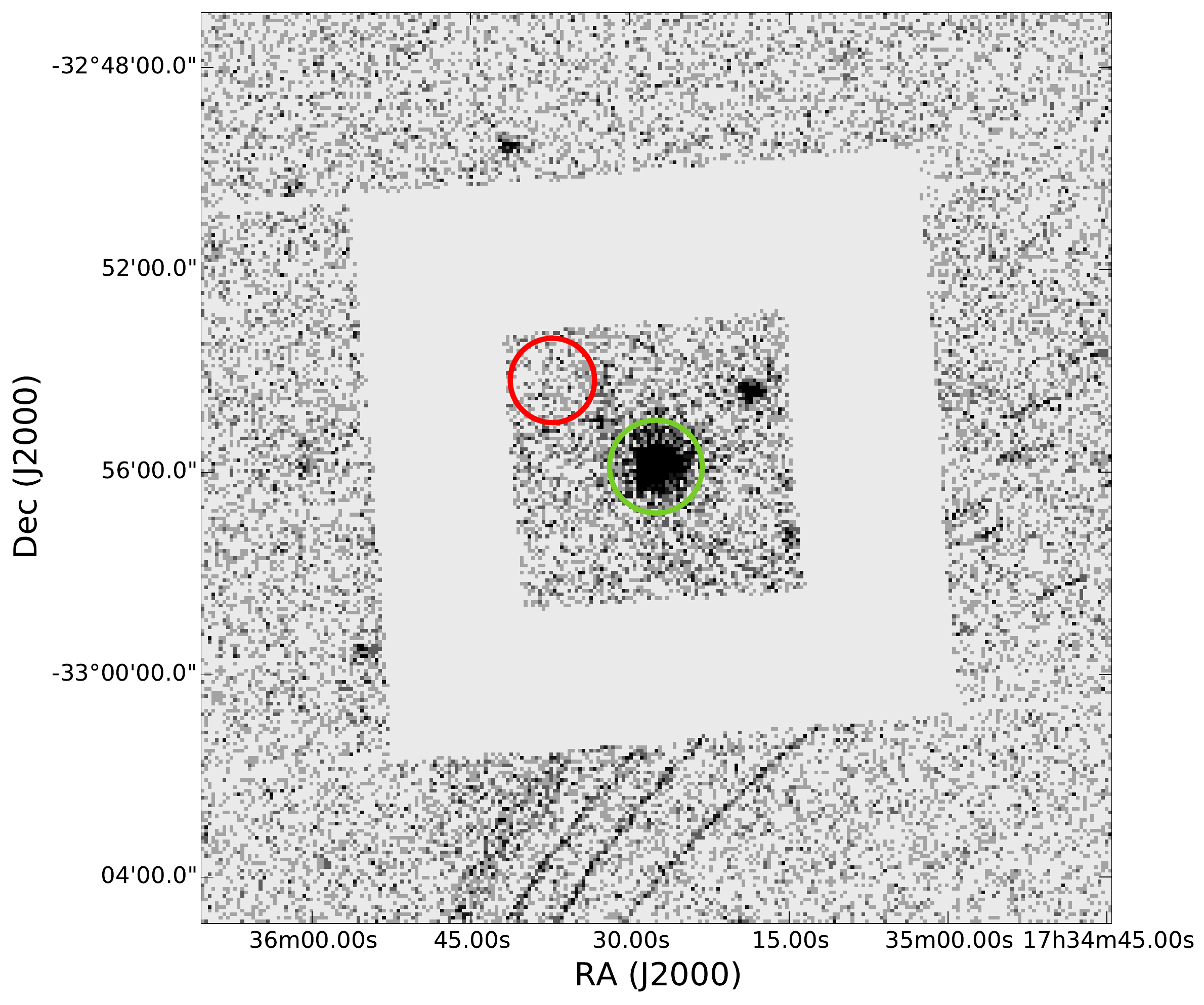}
	
\endminipage\hfill\newline
\minipage{0.4\textwidth}
	\centering
	\includegraphics[width=\columnwidth]{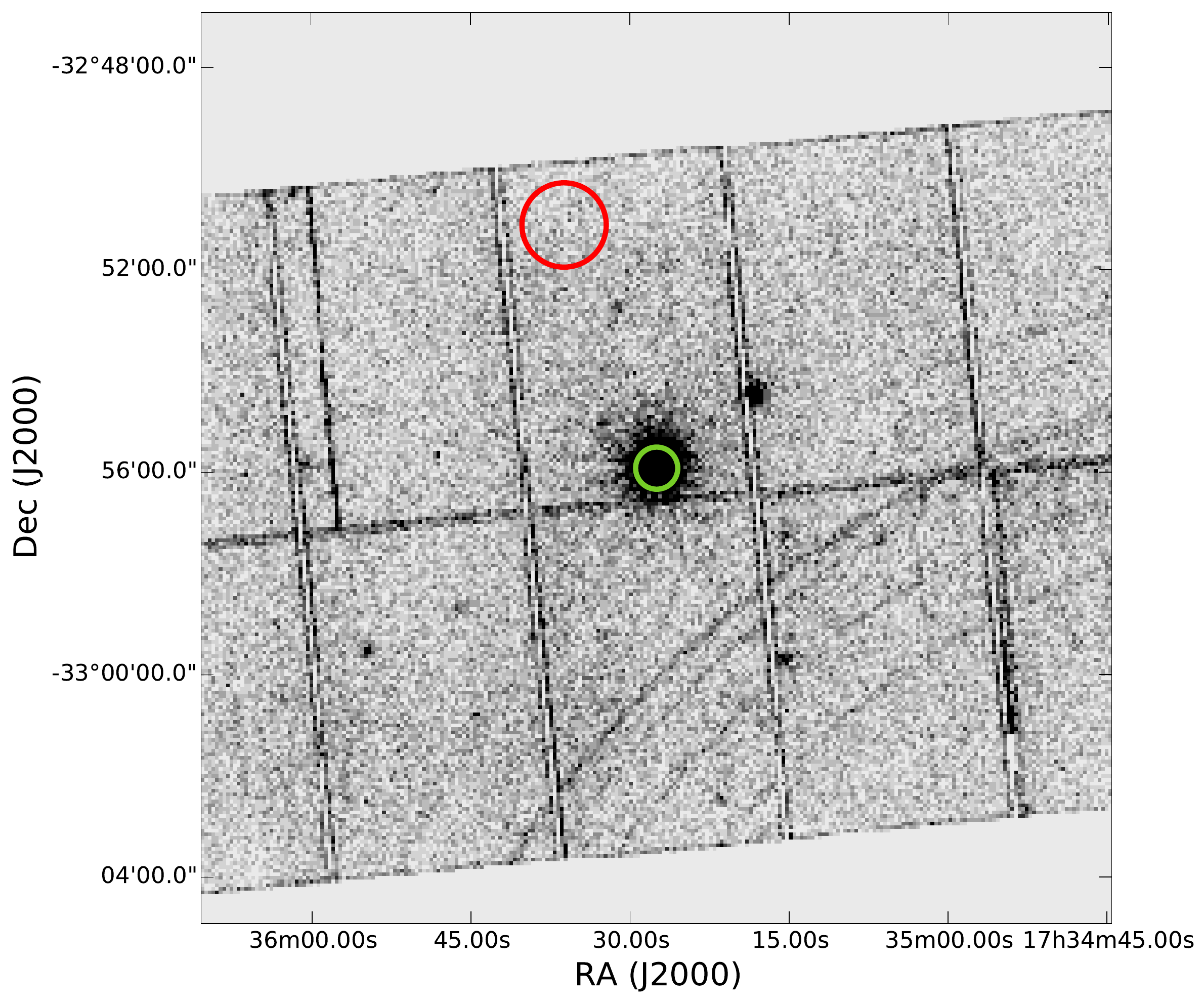}
	
\endminipage
\caption{EPIC-MOS1, EPIC-MOS2 and EPIC-pn images for observations one, two and three, respectively. The source and background extraction regions are shown as green and red circles, respectively. The excised inner source region in observation one is shown in yellow. \label{fig:images}}
\end{figure}

Source and background extraction regions were selected to optimise signal-to-noise using the {\scshape eregionanalyse} tool, but selections were subject to limitations such as the presence of chip gaps, and the changing observation configuration between observations one and two. The radius of each source region was chosen to be as wide as possible to encompass most of the source photons; the recommended value corresponds to an encircled energy fraction of $\sim80\%$ from the MOS and pn cameras according to the SAS users' handbook\footnote{https://xmm-tools.cosmos.esa.int/external/xmm\_user\_support/\\documentation/sas\_usg/USG/}. As can be seen in Figure \ref{fig:images}, however, the source was sometimes located close to the edge of a CCD so that the radius had to be restricted so as to avoid the collection of photons from adjacent CCDs.  In particular, the source region radius for the EPIC-pn camera in observation three had to be limited due to the source's extreme proximity to the chip gap. 

The background regions were nominally chosen to be about twice the size as the source regions subject to sufficient available space on the CCD (Figure \ref{fig:images}), but that was compromised heavily in the observations where the EPIC-MOS cameras were in large window mode, where the source and background regions were eventually of similar size.  Any differences in size of the source and background extraction areas were accounted for using the {\scshape backscale} and {\scshape lccorr} SAS tasks for the spectra and lightcurves, respectively. 

The maximum count rates of both the EPIC-MOS1 and EPIC-MOS2 cameras in observation one are slightly more than twice the tolerance limits for pile-up mentioned in \citet{Jethwa2015}, and this observation may therefore be affected by pile-up, likely due to operation in full frame mode. This is not the case for either of the cameras in the other two observations that were performed in large window mode. 
For completeness, all data sets were assessed for photon pile-up, using the {\scshape epatplot} task, which compares the observed-to-model fractions for single and double events. As expected based on the count rates, all single/double fractions were consistent with the nominal value of 1.0 within statistical errors (1 sigma) and there was good graphical agreement between the observed and expected pattern distributions. We conclude that observations two and three are not affected by pile-up. In both cameras for observation one the single/double event fractions are not always consistent with 1.0 implying that this observation may be affected by pile-up. In order to mitigate any pile-up, the core of the PSF was excised using an annular region around the centroid until the observed pattern distributions matched the expected ones and the single/double event fractions were consistent with 1.0. It was found that an annular inner radius of 200 physical pixels provided the best result for both EPIC-MOS1 and EPIC-MOS2 cameras. 

The resulting coordinates for both the source and background regions can be found in Table \ref{table:sourcebackground}.

All lightcurves and spectra reported in this paper were extracted following the standard procedures. The SAS tools {\scshape rmfgen} and {\scshape arfgen} were used to generate the response and ancillary matrix files. The files were grouped using the {\scshape specgroup} task with a minimum of 25 counts per bin and a maximum oversampling factor of three. 
Spectra were fitted with models using XSPEC version 12.8.2 where the uncertainties quoted are at the $90\%$ confidence level and the elemental abundances are set to those of \cite{Wilms2000}.

\section{analysis and results}
\subsection{\INTEGRAL}
\subsubsection{IBIS/ISGRI}
All available science windows covering the period from 2013 March 26 05:13 UTC to 2013 March 28 17:38 UTC for revolution 1276 and 2013 March 29 04:55 UTC to 2013 March 31 19:56 UTC for revolution 1277 were analysed. 
The source was not detected by IBIS/ISGRI in either energy bands (18$-$60 keV or 60$-$100 keV) in each single science window, in each single revolution significance map (1276, 1277) or in the summed revolution significance map 1276+1277. In the latter case (329 ks exposure) the 3$\sigma$ upper limit on the persistent flux is 0.77 mCrab or $1\times10^{-11}$\fluxunits (18$-$60 keV) while the 3$\sigma$ upper limit from the single revolution 1277 around the periastron passage (143 ks exposure) is 1.04 mCrab or $1.35\times10^{-11}$\fluxunits (18$-$60 keV).

\subsubsection{JEM-X}
We analysed all available JEM$-$X data acquired simultaneously with IBIS/ISGRI data during revolution 1277.
IGR J17354$-$3255 was in the JEM-X fully coded FOV at all times, however, it was not detected in a combined mosaic of JEM-X1 and JEM-X2 ($185$ ks combined exposure) in either the 3-10 keV or 10-20 keV energy bands. The $3\sigma$ upper flux limit was found to be $1.35\times10^{-11}$\fluxunits  (0.9 mCrab) in the 3$-$10 keV band.

\subsection{\XMM}
The background-subtracted light curves of IGR J17354$-$3255 are shown in Figure \ref{fig:bayesian}. As previously mentioned there was no EPIC-pn data in the first observation for this source. The light curves display moderately bright X-ray flares but there are no particularly strong outbursts in any of the three observations. The flux of the two flares in observation one $\sim4\times10^{-11}\mathrm{erg}$ $\mathrm{cm}^{-2}\mathrm{s}^{-1}$ is significantly higher than the average flux in observations two and three $\sim1.2-2.5\times10^{-11}\mathrm{erg}$ $\mathrm{cm}^{-2}\mathrm{s}^{-1}$. These two flares each lasted for $\sim7$ ks with a $\sim13$ ks gap between them. To date the strongest flare detected in the soft X-rays is the one detected by XRT ($8.7\times10^{-11}$\fluxunits \citep{Sguera2011}). The two flares detected by \XMM have a flux that is of the same order of magnitude as the XRT flare. The corresponding exposure times of the different cameras are outlined in Table \ref{table:one}.

\subsection{Spectral Variability}
We constructed light curves with a 300 s binning in soft (0.2$-$4 keV) and hard (4$-$10 keV) energy bands in order to investigate how the hardness ratio (HR) varies with time. This bin size was chosen (by inspection) as a compromise between having sufficient resolution to follow the rapid changes in source flux and retaining useful statistics within each bin.

\begin{equation}
	HR=\frac{H-S}{H+S}
\end{equation}

\begin{figure*}
	\begin{center}
		\includegraphics[width=\textwidth]{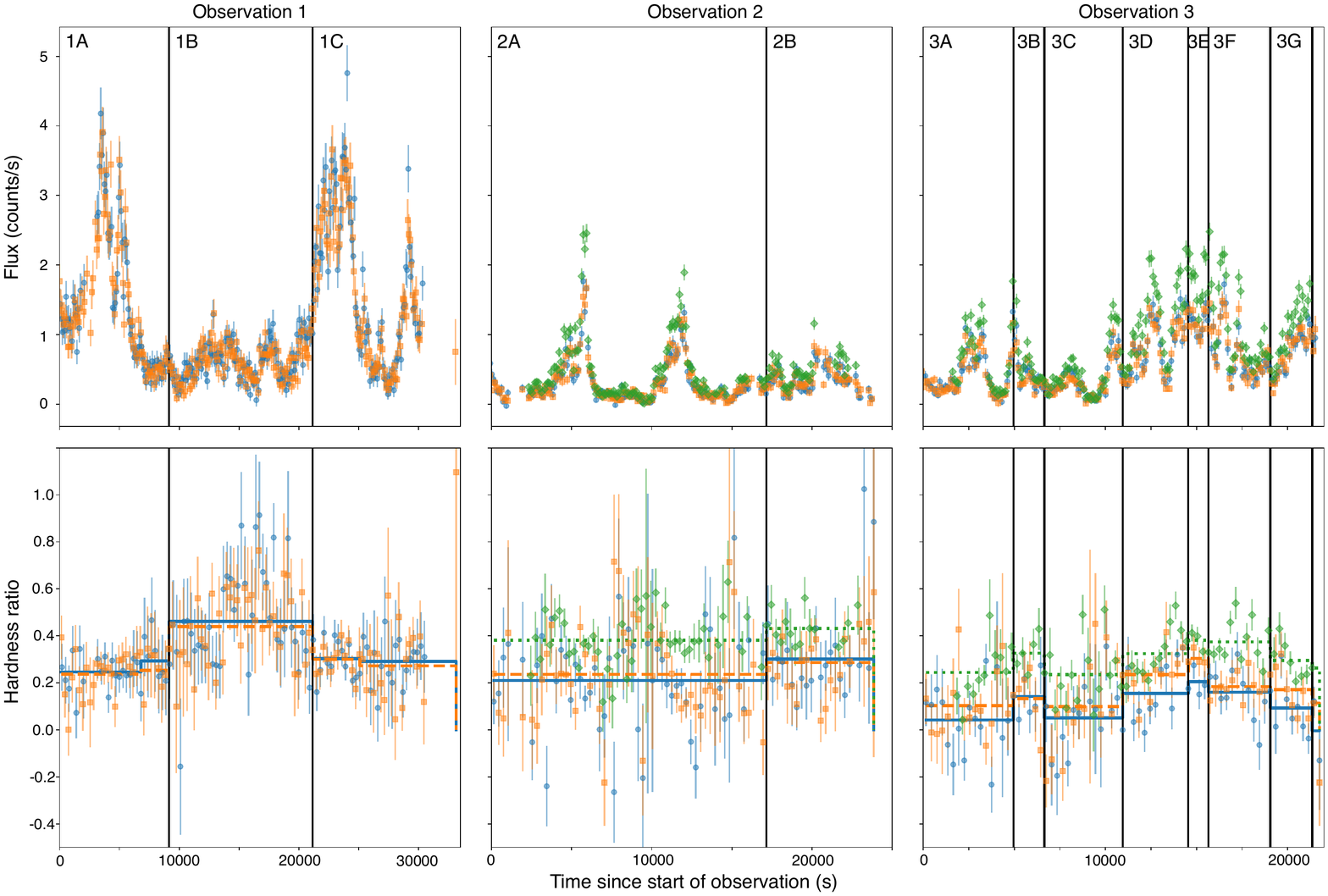}
		\caption{\label{fig:bayesian} Top panel: Light curves of the three observations in EPIC-MOS (blue and orange) and EPIC-pn (green) cameras. EPIC-pn fluxes are scaled by a factor of 0.5. Bottom panel: Bayesian block analysis of the hardness ratios for the three observations where the different blocks suggest the optimal segmentation of the data. The vertical lines represent the time stamps at which the observations were split into different sections, annotated as 1A-3G.}
	\end{center}
\end{figure*}

The hardness ratio light curve is shown in Figure \ref{fig:bayesian} (lower), and indicates that the spectral shape was changing during observation one, likely remained the same during observation two and changed more variably during observation three.
To objectively characterise any changes in hardness during the three observations we used the refined Bayesian blocks method of \cite{Scargle2013} to partition the dataset into sections with consistent hardness ratio. The purpose of the algorithm is to identify and characterise statistically significant variations using a simple non-parametric model that finds the optimal segmentation of the data without making any assumptions about the timescales or amplitude/shape of variations \citep{Hill2016}.
Additionally, the algorithm can perform multi-variate analysis so that the data from both the EPIC-MOS1 and EPIC-MOS2 cameras, as well as the EPIC-pn camera where available, could be modelled simultaneously (Figure \ref{fig:bayesian}). The small discrepancies between blocks, detected in different cameras, are due to differences in their spectral responses.

From this analysis it is clear that  the spectrum of IGR J17354$-$3255 hardens during a part of observation one extending from $\sim 9145-21150$ s from the start of the observation.
Consequently, we split the data from observation one into three distinct sections (labelled 1A, 1B and 1C) where the source spectrum transitions from being soft to harder and back to softer. When the source spectrum is softer the corresponding flux is greater, and vice versa. This may indicate that a dense clump of material is present in between the flares obscuring the source from view (see section \ref{discuss} Discussion). 

In the second observation, the source is in a state of low level activity. Small fluctuations can be seen in the light curve, however the Bayesian block analysis suggests only two segmentations from the hardness ratios. These are only minimally different so that we did not expect them to represent any significant spectral variations. For the sake of consistency we split the observation into two sections labelled 2A and 2B.

Inspection of the light curve for observation three suggests that IGR J17354$-$3255 was undergoing flaring activity on short time scales. Indeed the Bayesian block analysis suggests several segmentations when considering the hardness ratio. The differences between the sections are significant enough to warrant creating separate spectra for each; hence observation three was split into seven sections. There were insufficient counts in the final segment (from $\sim$21353s after the start time of the EPIC-MOS1 camera until the end of the observation) to create valid spectra. 

The observation subsections derived from the Bayesian blocks analysis and used in the subsequent spectral analysis are listed in Table~\ref{table:blocks}.

\begin{table}
\centering
\caption{Observation subsections derived from the Bayesian blocks analysis. Start and stop times are relative to the start of each observation.}
\label{table:blocks}
\begin{tabular}{cccccccccccc}
\hline
Section&Tstart&Tstop&Exposure\\
&[s]&[s]&[ks]\\
\hline
1A & 0       & 9150.0  & 9.2 \\
1B & 9150.0  & 21150.0 & 12.0 \\
1C & 21150.0 & 33100   & 12.0 \\
2A & 0       & 17152.6 & 17.2 \\
2B & 17152.6 & 24100   & 7.0 \\
3A & 0       & 4950.0  & 4.95 \\
3B & 4950.0  & 6653.0  & 1.73 \\
3C & 6653.0  & 10950.0 & 4.30 \\
3D & 10950.0 & 14550.8 & 3.60 \\
3E & 14550.8 & 15653.0 & 1.10 \\
3F & 15653.0 & 19050.0 & 3.40 \\
3G & 19050.0 & 21353.0 & 2.30 \\
\hline
\end{tabular}
\end{table} 
 
\subsection{Spectral Analysis}
Spectra were extracted for the individual sections and fitted with models using {\scshape XSPEC} version 12.8.2. The {\scshape cflux} model within {\scshape xspec} was used to give the unabsorbed flux of the entire model ({\scshape cflux} is a convolution model designed to determine the flux of a model or model component over a specified energy range).
A constant was used during simultaneous fitting in order to account for the difference between the EPIC-MOS and EPIC-pn cameras. Currently the instruments return energy-dependent flux differences of up to $7\%$ for broadband sources \citep{Read2014a}. Slight differences can occur between the two EPIC-MOS cameras due to different amounts of photons being received at the same time. The fit parameters resulting from all models are outlined in full in Tables \ref{table:powfits}, \ref{table:bbfits}, \ref{table:powgaussfits} and \ref{table:bbgaussfits}.

\subsubsection{Observation 1}

The first observation was divided into three sections based on the Bayesian block analysis of the hardness ratio evolution (Figure \ref{fig:bayesian}). For each of the sections the spectra obtained from the EPIC-MOS1 and EPIC-MOS2 cameras were fit simultaneously in the 1$-$12 keV energy range. A Tuebinger-Boulder ISM absoprtion model with a powerlaw {\scshape (const*tbabs*(powerlaw))} as well as one with a blackbody {\scshape (const*tbabs*(blackbody))} resulted in acceptable fits (Table \ref{table:powfits}). 
 
Inspection of the residuals for each section showed evidence of an emission feature consistent with a 6.4 keV iron-K$\alpha$ line. Consequently, to further characterise the spectral shape the spectrum was fit with an absorbed powerlaw or blackbody continuum with an additional Gaussian component {\scshape (const*tbabs*(powerlaw + gaussian)), (const*tbabs*(blackbody+gaussian))}. The absorption, powerlaw index, temperature, Gaussian line energy and line width were left as free parameters.

We found that the spectrum was well fit by this model in the case of an absorbed blackbody continuum (Table \ref{table:bbgaussfits}). When fitting an absorbed powerlaw not all sections had sensible parameters (Table \ref{table:powgaussfits}). 

While the powerlaw indices in sections 1A and 1C are consistent within errors it is apparent that the photon index in section 1B is slightly lower in comparison confirming that the source enters a harder spectral state during this time as was suggested by the HR analysis in section 3.1. A significant variation in hydrogen column density as part of the absorbed powerlaw model can also be seen with an $N_H$ of $\sim12 \times10^{22}\mathrm{cm}^{-2}$ in section 1A, increasing to $\sim17\times10^{22}\mathrm{cm}^{-2}$ in section 1B and finally decreasing again to $\sim14\times10^{22}\mathrm{cm}^{-2}$ in section 1C. The Gaussian line energy is consistent within errors between sections with an average value of $\sim6.38$ keV. 
 The Gaussian line widths correspond to  an electron plasma temperature of $\sim6\times10^{5}$ K which is well in excess of that expected for a supergiant atmosphere. This phenomenon can also be seen for some of the Gaussian line widths in the following two observations and is likely due to low signal-to-noise ratio. The EW of the line appears to be lower during the two flares compared to the gap between them suggesting that the iron emission is stronger between the two flares. The source was observed at an initial X-ray flux of $\sim3.8\times10^{-11}$ \fluxunits corresponding to an X-ray luminosity of $3.30\times10^{35}$ (assuming a distance of 8.5 kpc, as derived from the closeness to the Galactic centre and the high X-ray absorption; \citep{Tomsick2009}). There is a decrease in flux in section 1B and then again an increase in section 1C, coinciding with the previously mentioned differences in $\mathrm{N}_{H}$.
 Additionally, a small improvement in the residuals can be noted as the iron emission feature is introduced. These small improvements  suggest that there likely is an iron emission feature present.

\subsubsection{Observation 2}

The second observation was analysed in two sections as suggested by the Bayesian block analysis. 
The spectra extracted from the EPIC-MOS1, EPIC-MOS2 and EPIC-pn cameras were fitted simultaneously. We used the energy range 1$-$12 keV for the two EPIC-MOS cameras and 1-15 keV for the EPIC-pn camera.
An absorbed powerlaw model {\scshape (const*tbabs*(powerlaw))} resulted in acceptable fits with $\chi_{\mathrm{red}}^{2}/\mathrm{d.o.f}=1.23/324$ and $\chi_{\mathrm{red}}^{2}/\mathrm{d.o.f}=1.16/254$ (Table \ref{table:powfits}) for sections 2A and 2B, respectively. The hydrogen column density in section 2A is slightly lower than it was in the first observation whilst the one in section 2B is approximately the same as that from section 1A. The photon indices are consistent within errors with those of sections 1A and 1C respectively. The parameters compare similarly between the sections when using an absorbed blackbody model {\scshape (const*tbabs*(blackbody))}. The hydrogen column density is generally lower in the two sections in the second observation whilst the photon indices are consistent within errors with those of sections 1A and 1C. Considering there was evidence of a $\sim 6.4$ keV iron-K$\alpha$ line in observation one we inspected these spectra for the same feature. Consequently, we added a Gaussian component to the models {\scshape (const*tbabs*(powerlaw + gauss)), (const*tbabs*(blackbody+gauss)}. Unlike the first observation, the fits did not improve as much in this one and the iron lines are at different energies and do not have constrained values for the respective equivalent widths.
All fit parameters for this observation for the two different models are outlined in Tables \ref{table:powfits} to \ref{table:six}.

\begin{figure}
	\begin{center}
		\includegraphics[width=\columnwidth]{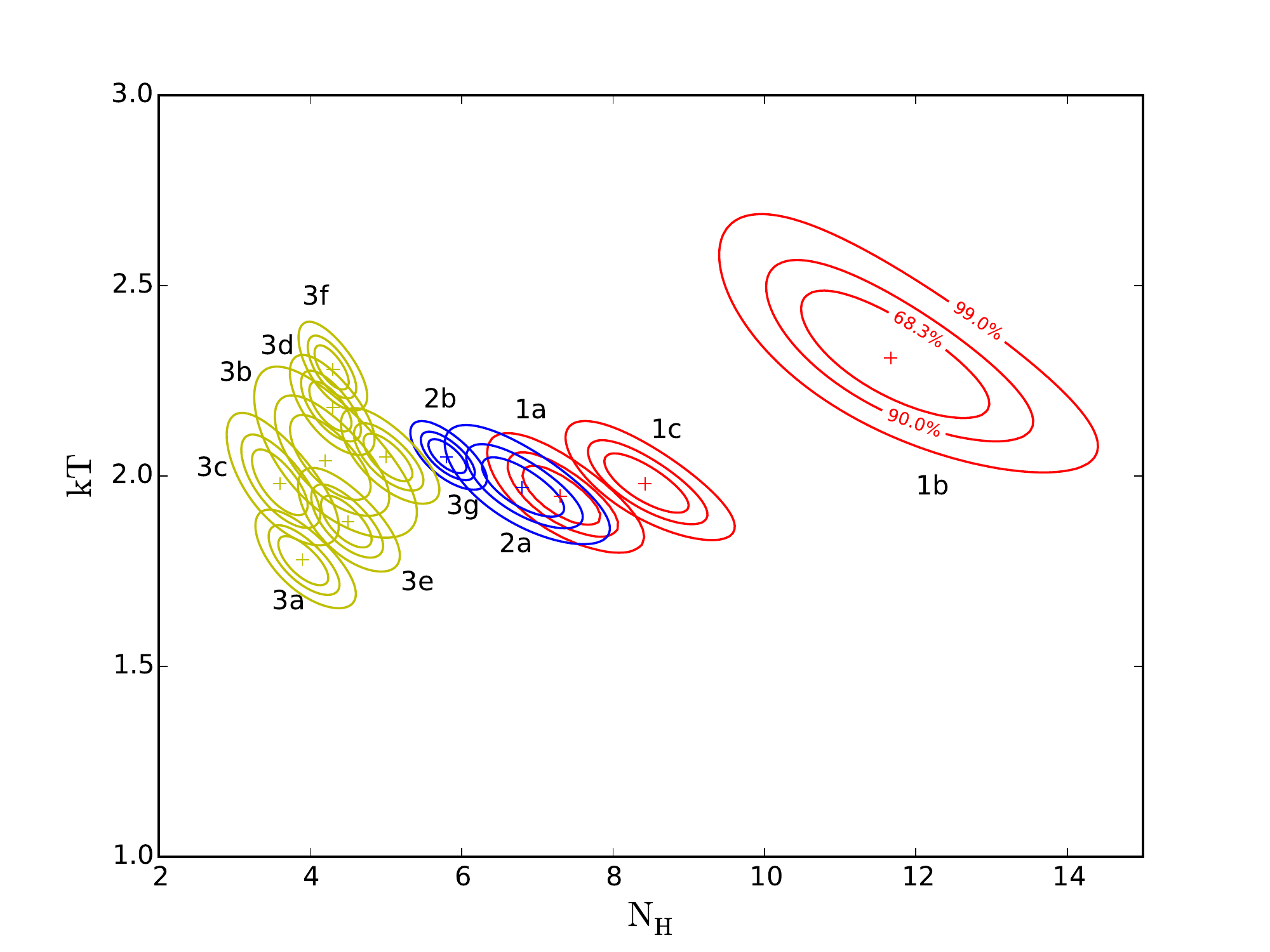}
		\caption{\label{fig:contour} Contour plots for the spectral parameters measured from all observations (Observation one is \textcolor{red}{red}, observation 2 is \textcolor{blue}{blue} and observation 3 is \textcolor{yellow}{yellow}). The contours correspond to the $68\%$, $90\%$ and $99\%$ confidence levels.}
	\end{center}
\end{figure}

\subsubsection{Observation 3}

Observation three was divided into seven sections. For each of the sections the spectra obtained from the EPIC-MOS1, EPIC-MOS2 and EPIC-pn cameras were fitted simultaneously using the same energy ranges as the previous two observations. At first an absorbed powerlaw model {\scshape (const*tbabs*(powerlaw))} was used resulting in mainly reasonable fitting parameters. The values for the absorption column density appear to be constant within errors except for section 3F where it is $9.7\pm0.3\times10^{22}\mathrm{cm}^{-2}$ instead of $\sim8\times10^{22}\mathrm{cm}^{-2}$. The photon indices are all around $1.1-1.3$ except the one in section 3A, which appears to be slightly higher with a value of 1.47. Similarly, the absorption column densities from the blackbody continuum appear to be constant within errors. We note that 2B and 3A are 2 days apart so given the variability observed, we have no expectation of continuity of parameter values.

Since, when inspecting the individual spectra a hint of a $\sim 6.4$ keV iron-K$\alpha$ line could be seen, the model was again updated to include this feature. As with the first observation the model {\scshape (const*tbabs*(powerlaw+gaussian))} resulted in moderately improved fits in those sections where the iron line was found (3B, 3F and 3G). When trying to fit this model to the other sections, however, the resulting iron line energy was not sensible. Setting the initial value of the parameter around an energy of 6.4 keV did not improve matters. Even so, the other parameters are still consistent with each other in the same way as they were when using an absorbed powerlaw model without a Gaussian component. 

The variation between some of the parameters as well as the $\chi_{\mathrm{red}}^{2}/\mathrm{d.o.f}$ values are likely due to the rapid fluctuations on short time scales for this observation. Whilst the Bayesian block analysis suggests to divide the data into seven individual sections some of them still show significant changes in flux occurring at short time scales, which can affect the quality of the extracted spectrum. A time-resolved spectral analysis with a 1000 s resolution using an absorbed powerlaw model (Table \ref{table:six}) resulted in $\chi_{\mathrm{red}}^{2}/\mathrm{d.o.f}$ values between $0.97/72$ and $1.47/182$. The $N_H$ stayed approximately constant with values $\sim8\times10^{22}\mathrm{cm}^{-2}$; solely towards the end of the observation does it once appear to be higher with a value of $10.2\times10^{22}\mathrm{cm}^{-2}$. The photon index shows small changes with values between 0.89 and 1.57. Overall, the individual spectra do not appear to be too dissimilar such that no different model is needed to interpret the spectral variation which is in agreement with the analysis of the seven sections resulting from the Bayesian block analysis previously.  Similarly, the absorbed blackbody model has hydrogen column density values of $\sim4\times10^{22}\mathrm{cm}^{-2}$ and temperatures between 1.72 and 2.44 kT. 

Much the same as in observation two the addition of the Gaussian component to model the iron emission feature is not statistically needed to interpret the different sections of observation three.  The observed X-ray flux varies throughout the observation, however it is never comparable to the flux of the two flares in observation one ranging from $\sim1.1-3.3\times10^{-11}$ \fluxunits. This corresponds to X-ray luminosities between $0.9-2.9\times10^{35}$ \lumunits. All fit parameters for both models are outlined in Tables \ref{table:powfits} to \ref{table:timeres}. 

Figure \ref{fig:contour} plots the spectral parameters obtained for all observations when consistently modelled with an absorbed blackbody continuum. It is evident that the contours overlap within each observation. One of the contours in observation two overlaps with one from observation 1 whilst the other one from observation two overlaps with one from observation 3. Consequently, a decrease in hydrogen column density can be observed throughout the observations. The only contour that is not connected to any other is the one from section 1B where the hydrogen column density is about twice as high as the average value obtained from the other sections.

Fit results for an absorbed powerlaw model and an absorbed blackbody model have been discussed in detail, and have been used primarily as consistent reference models when subdividing the dataset in a search for spectral variations. More complex, science-driven models often used in fitting SFXT spectra were also investigated, but in most cases failed due to a lack of statistics when considering short sections of data. For example, a Comptonizing plasma model {\scshape const*tbabs*(comptt)} and a cut-off power-law {\scshape const*tbabs*(cutoffpl)} failed to give better fits to the data and often resulted in unphysical or unconstrained parameters. Fitting a Comptonizing plasma model to section 1A results in \reducedchi= $1.43/173$ with kT=$215\pm2944$ keV, a seed photon temperature of $\mathrm{T}_{0}=0.3\pm1.0$ keV and an optical depth of $\tau=0.8\pm102.0$. As another example, a cut off power-law fitted to section 3A gave an $\mathrm{N}_{\mathrm{H}}=6.5\pm0.7\times10^{22}\mathrm{cm}^{-2}$, a photon index of $\Gamma=0.5\pm0.4$ and an high energy cut off of $6\pm2$ keV. These problems are likely due to the lack of significant data above 10 keV required to constrain the additional fit components.

\begin{figure}
	\begin{center}
		\includegraphics[width=\columnwidth]{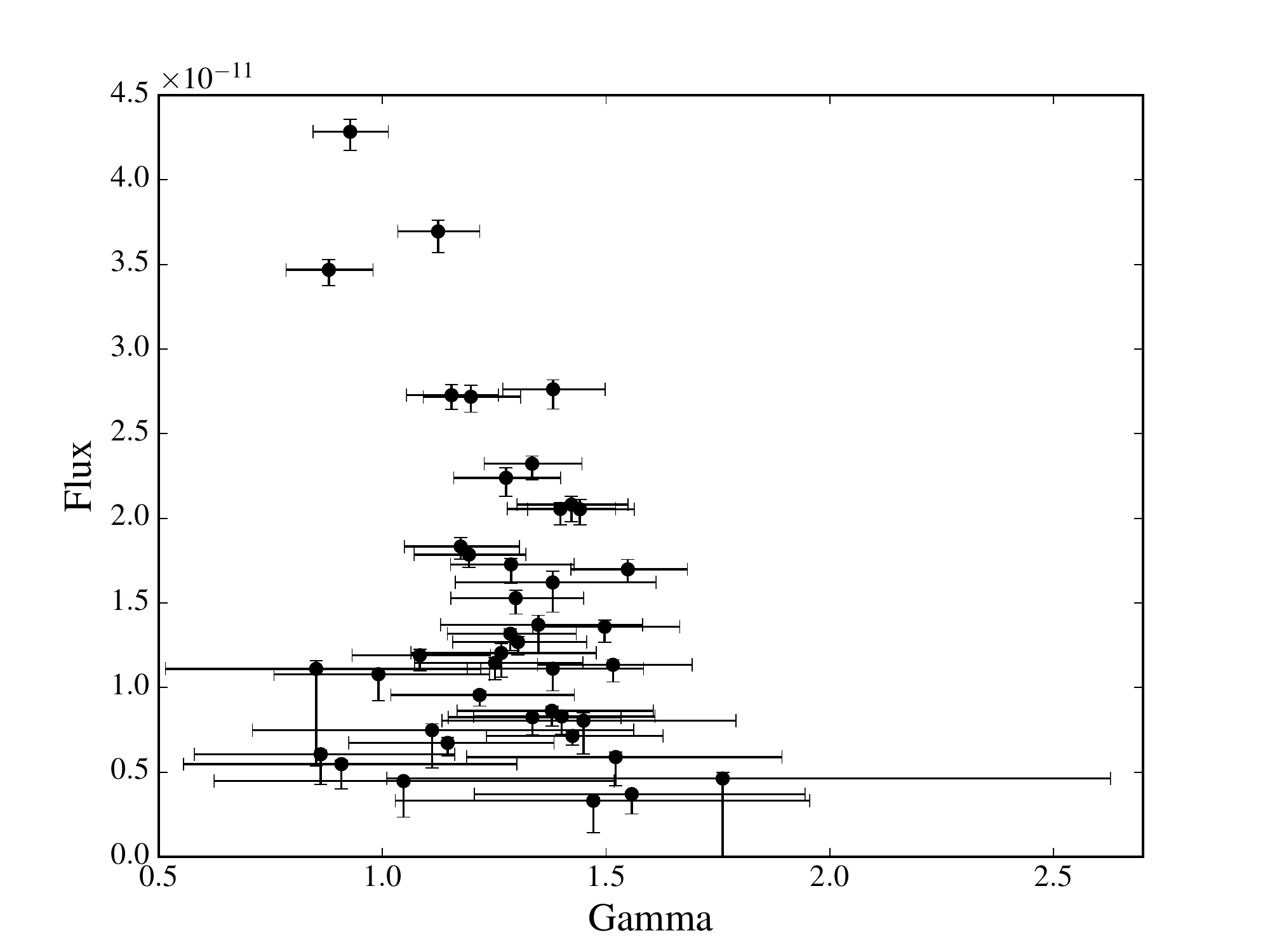}
		\caption{\label{fig:correlation} Time-resolved analysis of flux and photon index through observations two and three. Each data point is derived from a 1000s subsection of data.}
	\end{center}
\end{figure}

By performing time-resolved spectral analysis (using a simple powerlaw model) we investigated if there is a correlation between the hardness and flux of the source in observations two and three. The Pearson correlation coefficient (where a value of zero means there is no correlation, -1 is an exact negative, +1 is an exact positive linear correlation) was calculated to measure the linear relationship between hardness and flux. Combining the results of observations two and three resulted in a correlation coefficient of -0.27, implying that the flux values and photon indices are not correlated (Figure \ref{fig:correlation}). The same analysis was unsuccessful for observation one due to the higher error values, resulting from the pile-up reduction, making it impossible to create the individual spectra needed for the time-resolved spectral analysis.

\begin{table*}
\begin{minipage}{170mm}
\centering
\caption{Spectral parameters for each sub-observation obtained using a {\scshape tbabs*(powerlaw)} model. Uncertainties quoted are at the $90\%$ confidence level.}
\label{table:powfits}
%\resizebox{\columnwidth}{!}{
\begin{tabular}{ccccccccc}
%\hline
%\hline
%c*tbabs*(pow)&&&&&&&&&&&\\
\hline
Section&$\mathrm{N}_{\mathrm{H}}$&$\Gamma$&Unabsorbed Flux&Luminosity&$\chi_{\mathrm{red}}^{2}$/d.o.f\\
&$[\times10^{22}\mathrm{cm}^{-2}]$&&[$\times10^{-11}$erg $\mathrm{cm}^{-2}\mathrm{s}^{-1}$]&$[\times10^{35}$\lumunits]\\
\hline
1A & $12.1\pm0.5$&$1.26\pm0.07$&$5.5\pm0.2$&4.77&$1.42/175$\\
1B & $17\pm1$&$1.0\pm0.1$&$2.8\pm0.2$&2.46&$1.23/132$\\
1C & $14.1\pm0.5$&$1.30\pm0.07$&$5.9\pm0.2$&5.06&$1.43/199$\\
2A & $10.8\pm0.3$&$1.27\pm0.04$&$1.12\pm0.03$&0.90&$1.23/324$\\
2B & $12.7\pm0.6$&$1.42\pm0.07$&$1.18\pm0.06$&1.02&$1.16/254$\\
3A & $8.1\pm0.4$&$1.47\pm0.07$&$1.23\pm0.05$&1.07&$1.17/206$\\
3B & $8.2\pm0.5$&$1.16\pm0.09$&$1.8\pm0.1$&1.53&$1.59/117$\\
3C & $7.4\pm0.4$&$1.19\pm0.07$&$1.09\pm0.06$&0.94&$1.14/178$\\
3D & $8.4\pm0.3$&$1.07\pm0.05$&$2.8\pm0.1$&2.43&$1.32/254$\\
3E & $8.3\pm0.2$&$0.97\pm0.04$&$3.3\pm0.1$&2.85&$1.27/298$\\
3F & $9.7\pm0.3$&$1.23\pm0.05$&$2.6\pm0.1$&2.27&$1.38/247$\\
3G & $8.5\pm0.4$&$1.34\pm0.06$&$2.4\pm0.1$&$2.06$&$1.08/199$\\
\hline
\end{tabular}%}
\end{minipage}
\end{table*} 

\begin{table*}
\begin{minipage}{170mm}
\centering
\caption{Spectral parameters for each sub-observation obtained using a {\scshape tbabs*(blackbody)} model. Uncertainties quoted are at the $90\%$ confidence level.}
\label{table:bbfits}
%\resizebox{\columnwidth}{!}{
\begin{tabular}{ccccccccc}
%\hline
%\hline
%c*tbabs*(bb)&&&&&&&&&&&\\
\hline
Section&$\mathrm{N}_{\mathrm{H}}$&$kT$&Unabsorbed Flux&Luminosity&$\chi_{\mathrm{red}}^{2}$/d.o.f\\
&$[\times10^{22}\mathrm{cm}^{-2}]$&[keV]&[$\times10^{-11}$erg $\mathrm{cm}^{-2}\mathrm{s}^{-1}$]&$[\times10^{35}$\lumunits]\\
\hline
1A & $7.3\pm0.3$&$1.95\pm0.05$&$5.2\pm0.2$&4.52&$1.21/179$\\
1B & $11.7\pm0.8$&$2.3\pm0.1$&$2.7\pm0.2$&2.35&$1.13/132$\\
1C & $8.4\pm0.3$&$1.98\pm0.05$&$5.6\pm0.2$&4.82&$1.31/199$\\
2A & $6.8\pm0.3$&$1.97\pm0.05$&$1.16\pm0.06$&1.00&$1.18/254$\\
2B & $5.8\pm0.2$&$2.05\pm0.03$&$1.09\pm0.03$&0.95&$1.36/324$\\
3A & $3.9\pm0.2$&$1.78\pm0.04$&$1.19\pm0.05$&1.03&$1.31/206$\\
3B & $4.2\pm0.4$&$2.04\pm0.07$&$1.7\pm0.1$&1.50&$1.55/117$\\
3C & $3.6\pm0.2$&$1.98\pm0.05$&$1.04\pm0.06$&0.90&$1.25/178$\\
3D & $4.3\pm0.1$&$2.18\pm0.04$&$2.8\pm0.1$&2.38&$1.30/254$\\
3E & $4.3\pm0.1$&$2.28\pm0.04$&$3.2\pm0.1$&2.81&$1.27/298$\\
3F & $5.0\pm0.2$&$2.05\pm0.04$&$2.6\pm0.1$&2.22&$1.40/247$\\
3G & $4.5\pm0.2$&$1.88\pm0.04$&$2.3\pm0.1$&$1.98$&$1.18/199$\\
\hline
\end{tabular}%}
\end{minipage}
\end{table*} 

\begin{table*}
\begin{minipage}{170mm}
\centering
\caption{Spectral parameters for each sub-observation obtained using a {\scshape tbabs*(powerlaw+gaussian)} model. Uncertainties quoted are at the $90\%$ confidence level.}
\label{table:powgaussfits}
%\resizebox{\columnwidth}{!}{
\begin{tabular}{ccccccccccccc}
%\hline
%\hline
%c*tbabs*(pow+gauss)&&&&&&&&&&\\
\hline
Section&$\mathrm{N}_{\mathrm{H}}$&$\Gamma$&lineE&$EW_{\mathrm{line}}$&Unabsorbed Flux&Luminosity&$\chi_{\mathrm{red}}^{2}$/d.o.f\\
&$[\times10^{22}\mathrm{cm}^{-2}]$&&[keV]&[keV]&[$\times10^{-11}$erg $\mathrm{cm}^{-2}\mathrm{s}^{-1}$]&$[\times10^{35}$\lumunits]\\
\hline
1A & $12.2\pm0.5$ &$1.30\pm0.07$ & $6.36\pm0.04$ & $0.10\pm0.06$ & $5.5\pm0.2$   & 4.74 & $1.36/172$ \\
1B & $17\pm1$     &$1.0 \pm0.1$  & $6.32\pm0.06$ & $0.13\pm0.08$ & $2.8\pm0.2$   & 2.43 & $1.16/129$ \\
1C & $15.0\pm0.3$ &$1.48\pm0.05$ & $6.38\pm0.2$  & $0.05\pm0.04$          & $4.02\pm0.09$ & 3.47 & $1.13/242$ \\
2A & $10.8\pm0.3$ &$1.28\pm0.04$ & $6.45\pm0.05$ & $0.03\pm0.02$             & $1.12\pm0.03$ & 0.97 & $1.21/321$ \\
2B & $12.7\pm0.6$ &$1.44\pm0.07$ & $6.87\pm0.08$ & $0.07\pm0.06$             & $1.18\pm0.06$ & 1.02 & $1.15/251$ \\
3A &  $8.1\pm0.4$ &$1.47\pm0.07$ & $6.5\pm0.01$ &$0.06\pm0.04$             & $1.24\pm0.06$ & 1.07 & $1.15/204$ \\
3B &  $8.2\pm0.5$ &$1.19\pm0.09$ & $6.4\pm0.3$   & $0.14\pm0.07$         & $1.8\pm0.1$   & 1.52 & $1.53/114$ \\
3C &  $7.4\pm0.4$ &$1.20\pm0.07$ & $6.06\pm0.09$ & $0.05\pm0.04$          &$1.08\pm0.06$       & 0.9  & $1.13/176$ \\
3D &  $8.4\pm0.3$ &$1.08\pm0.05$ & $6.43\pm0.05$ & $0.04\pm0.03$ & $2.8\pm0.1$      & 2.43 & $1.30/252$ \\
3E &  $8.3\pm0.2$ &$0.98\pm0.04$ & $6.43\pm0.04$ & $0.04\pm0.02$ &$3.3\pm0.1$& 2.84 & $1.25/296$ \\
3F &  $9.6\pm0.3$ &$1.25\pm0.05$ & $6.36\pm0.02$ & $0.10\pm0.04$ & $2.6\pm0.1$   & 2.26 & $1.29/244$ \\
3G &  $8.6\pm0.4$ &$1.37\pm0.07$ & $6.39\pm0.05$ & $0.12\pm0.06$ & $2.4\pm0.1$   & 2.06 & $1.03/196$ \\
\hline
\end{tabular}%}
\end{minipage}
\end{table*}

\begin{table*}
\begin{minipage}{170mm}
\centering
\caption{Spectral parameters for each sub-observation obtained using a {\scshape tbabs*(blackbody+gaussian)} model. Uncertainties quoted are at the $90\%$ confidence level.}
\label{table:bbgaussfits}
%\resizebox{\columnwidth}{!}{
\begin{tabular}{ccccccccccccc}
%\hline
%\hline
%c*tbabs*(bb+gauss)&&&&&&&&&&\\
\hline
Section&$\mathrm{N}_{\mathrm{H}}$&$kT$&lineE&$EW_{\mathrm{line}}$&Unabsorbed Flux&Luminosity&$\chi_{\mathrm{red}}^{2}$/d.o.f\\
&$[\times10^{22}\mathrm{cm}^{-2}]$&[keV]&[keV]&[keV]&[$\times10^{-11}$erg $\mathrm{cm}^{-2}\mathrm{s}^{-1}$]&$[\times10^{35}$\lumunits]\\
\hline
1A&$7.3\pm0.3$&$1.92\pm0.05$&$6.4\pm0.05$&$0.08\pm0.04$&$5.1\pm0.2$&4.45&$1.16/179$\\
1B&$11.5\pm0.8$&$2.3\pm0.1$&$6.33\pm0.06$&$0.11\pm0.07$&$2.6\pm0.1$&2.2&$1.08/129$\\
1C&$8.4\pm0.4$&$1.96\pm0.05$&$6.40\pm0.04$&$0.05\pm0.05$&$5.6\pm0.2$&4.80&$1.29/203$\\
2A&$5.8\pm0.2$&$2.05\pm0.03$&$6.4\pm0.1$&$0.01\pm0.01$&$1.09\pm0.04$&0.95&$1.37/321$\\
2B&$6.8\pm0.3$&$1.97\pm0.05$&$6.9\pm0.3$&$0.03\pm0.03$&$1.16\pm0.06$&1.00&$1.20/251$\\
3A&$3.9\pm0.2$&$1.77\pm0.04$&$6.4\pm0.1$&$0.04\pm0.02$&$1.19\pm0.06$&1.03&$1.32/203$\\
3B&$4.2\pm0.4$&$2.02\pm0.07$&$6.4\pm0.1$&$0.10\pm0.06$&$1.7\pm0.1$&1.49&$1.52/114$\\
3C&$3.6\pm0.2$&$1.98\pm0.06$&$6.8\pm0.5$&$0.01\pm0.03$&$1.7\pm0.1$&0.90&$1.27/175$\\
3D&$4.2\pm0.2$&$2.18\pm0.04$&$6.4\pm0.4$&$0.01\pm0.02$&$2.8\pm0.1$&2.38&$1.30/251$\\
3E&$4.3\pm0.1$&$2.28\pm0.04$&$6.4\pm0.2$&$0.02\pm0.20$&$3.2\pm0.1$&2.80&$1.28/295$\\
3F&$5.0\pm0.2$&$2.03\pm0.04$&$6.36\pm0.07$&$0.07\pm0.02$&$2.6\pm0.1$&2.21&$1.35/244$\\
3G&$4.5\pm0.2$&$1.86\pm0.04$&$6.40\pm0.07$&$0.07\pm0.06$&$2.3\pm0.1$&2.21&$1.17/196$\\
\hline
\end{tabular}%}
\end{minipage}
\end{table*} 

\begin{table*}
\begin{minipage}{170mm}
\centering
\caption{Time-resolved spectral analysis of observation three with a resolution of 1000s using the {\scshape tbabs*blackbody} model. Uncertainties quoted are at the $90\%$ confidence level.}
\label{table:timeres}
\begin{tabular}{llllll}
%\hline
\hline
Number &$\chi_{\mathrm{red}}^{2}$& dof & $\mathrm{N}_{\mathrm{H}}$ & kT & Flux \\
&&&$[\times10^{22}\mathrm{cm}^{-2}]$&[keV]&[$\times10^{-11}$erg $\mathrm{cm}^{-2}\mathrm{s}^{-1}$]\\
\hline
\hline
1      & 1.12   & 68  & 4.0 $\pm$ 1.1 &   1.7 $\pm$ 0.2 &   0.74 $\pm$ 0.05     \\
2      & 1.11   & 180 & 3.5 $\pm$ 0.5 &   1.7 $\pm$ 0.1 &   1.29 $\pm$ 0.06   \\
3      & 1.04   & 214 & 3.8 $\pm$ 0.4 &   1.7 $\pm$ 0.1 &   1.63 $\pm$ 0.06   \\
4      & 1.00   & 168 & 4.4 $\pm$ 0.7 &   1.8 $\pm$ 0.1 &   1.09 $\pm$ 0.05   \\
5      & 1.49   & 178 & 4.8 $\pm$ 0.6 &   1.9 $\pm$ 0.1 &   1.23 $\pm$ 0.05   \\
6      & 1.39   & 185 & 4.3 $\pm$ 0.5 &   1.9 $\pm$ 0.1 &   1.28 $\pm$ 0.04   \\
7      & 1.19   & 122 & 3.6 $\pm$ 0.7 &   1.8 $\pm$ 0.1 &   0.69 $\pm$ 0.04  \\
8      & 1.08   & 134 & 3.4 $\pm$ 0.7 &   1.9 $\pm$ 0.1 &   0.79 $\pm$ 0.04   \\
9      & 1.08   & 106 & 3.5 $\pm$ 0.8 &   2.0 $\pm$ 0.2 &   0.65 $\pm$ 0.04    \\
10     & 0.99   & 166 & 3.7 $\pm$ 0.6 &   2.1 $\pm$ 0.1 &   1.15 $\pm$ 0.05   \\
11     & 1.04   & 218 & 3.9 $\pm$ 0.4 &   2.0 $\pm$ 0.1 &   1.74 $\pm$ 0.05   \\
12     & 1.04   & 237 & 4.2 $\pm$ 0.4 &   1.9 $\pm$ 0.1 &   2.27 $\pm$ 0.07   \\
13     & 1.03   & 251 & 4.3 $\pm$ 0.4 &   2.1 $\pm$ 0.1 &   2.69 $\pm$ 0.08   \\
14     & 1.13   & 269 & 4.9 $\pm$ 0.4 &   2.4 $\pm$ 0.1 &   3.44 $\pm$ 0.08   \\
15     & 1.26   & 299 & 4.6 $\pm$ 0.3 &   2.4 $\pm$ 0.1 &   4.23 $\pm$ 0.09   \\
16     & 1.16   & 281 & 4.7 $\pm$ 0.4 &   2.2 $\pm$ 0.1 &   3.63 $\pm$ 0.08 \\
17     & 1.17   & 249 & 5.1 $\pm$ 0.5 &   2.1 $\pm$ 0.1 &   2.66 $\pm$ 0.08  \\
18     & 1.05   & 213 & 5.0 $\pm$ 0.5 &   2.1 $\pm$ 0.1 &   1.80 $\pm$ 0.06   \\
19     & 0.98   & 191 & 5.3 $\pm$ 0.6 &   2.0 $\pm$ 0.1 &   1.49 $\pm$ 0.05   \\
20     & 1.08   & 237 & 4.8 $\pm$ 0.4 &   1.8 $\pm$ 0.1 &   1.99 $\pm$ 0.07 \\
\hline
\end{tabular}%}
\end{minipage}
\end{table*}

\subsection{Temporal Analysis}
Less than half of the confirmed SFXTs have a known spin period for their suspected neutron star accretor. Those identified range from a possible $\sim5\mathrm{s}$ in AX J1841.0$-$0536 \citep{Bamba2001} to $\sim1212\mathrm{s}$ in IGR J16418$-$4532 \citep{Sidoli2012a}. Previous observations of IGR J17354$-$3255 have failed to find a periodic signal that can be interpreted as the spin period of the neutron star \citep{Ducci2013}. 

We searched for periodicities in the range 0.1 -- 500s using both unbinned and binned (where appropriate) methods. The primary method used was a $Z^2_n$ method \citep{z-squared} applied to event lists created from the source locations in the cleaned observation event lists. The upper period range for which our search was set by the timescales of the flaring structures ($\sim$1000s) in the light curves, which inject red noise into the analysis. These tests were supported by randomisation monte-carlo tests to determine the noise levels within the periodiograms, but we must consider that such tests are only sensitive to white noise, and provide only an lower limit to the confidence levels for powers seen in the periodograms.

We found no evidence for a persistent periodicity in the light curves for Observations 1 and 2, the periodogram powers seen up to $\sim$500s is consistent with white noise, and beyond 500s the periodicity analysis is strongly impacted by the non-periodic structures easily seen in the light curves on ~1000s timescales. There was tentative evidence for a 251s period in the three light curves of Obs3, but in this case the source PSF is badly affected by a CCD feature which cannot be corrected for in an event-based analysis.

Binned light curves were extracted for all cameras in observation 1 (10s binning) and observations 2 and 3 (1s binning); the bin sizes were chosen as a compromise between avoiding empty bins considering the mean count rates of the source, and retaining sensitivity to short-period signals. As a result of this necessary compromise, binned methods are less optimal for low statistics light curves such as these. A Lomb-Scargle periodogram (\citet{Lomb1976}; \citet{Scargle1982}) was produced using the fast implementation of \citet{Press1989} for each binned light curve. These tests were again supported by randomisation monte-carlo tests to determine the noise levels within the periodiograms. No significant ($>99\%$) probability periodicities were detected in the range 2$x$binsize -- 500s. An epoch folding method using the $\mathrm{Q}^{2}$ statistic as defined by \citet{Leahy1983} was used to further test the binned light curves for period signals. Again, no significant peaks were identified. The binned analysis was unable to replicate or confirm the tentative 251s period seen in the unbinned analysis.

%% discussion

\section{Discussion}\label{discuss}

We present the results of three \XMM observations of the SFXT, IGR J17354$-$3255. Observation one was performed at a phase range $\phi=0.154-0.199$, as an initial trial pointing, resulting in the source only being observed by the two EPIC-MOS cameras. Observations two and three were carried out approximately two weeks later closer to system periastron (at phase ranges $\phi=0.804-0.839$ and $\phi=0.996-0.025$ respectively) and with the instruments in a more appropriate configuration. Significant flux variations and flaring were detected in all three observations. In order to avoid averaging spectral analysis across different source behaviours, a Bayesian block method was utilised in order to objectively characterise possible changes in spectral hardness through the three observations. As a result, the three observations were split into twelve separate sub-sections for which individual spectra were extracted. 

\subsection{Continuum emission and absorption column}

A spectral analysis of all the sections revealed that the source continuum X-ray emission could be described just as well by either a Tuebinger-Boulder ISM absorption model with a powerlaw or a blackbody.

The photon index derived from the power-law fits varied over the range $\Gamma=\sim1.0-1.5$, with a typical error of $\sim\pm0.1$, suggesting significant changes. However, there is no obvious trend or evolution indicating that these changes in photon index are correlated with other parameters (see later).

The absorption column density varied significantly across the three sections of observation one, reaching $\mathrm{N}_{\mathrm{H}}\sim17\times10^{22}\mathrm{cm}^{-2}$ at its highest. This corresponded in time to a period of low flux between flaring. The power-law fits for observation two and observation three showed slowly declining $\mathrm{N}_{\mathrm{H}}$ values of typically  $\mathrm{N}_{\mathrm{H}}\sim11\times10^{22}\mathrm{cm}^{-2}$ and  $\mathrm{N}_{\mathrm{H}}\sim8\times10^{22}\mathrm{cm}^{-2}$ respectively. The same trend was seen in the blackbody model fits, where the absorption column density increased to a maximum of  $\mathrm{N}_{\mathrm{H}}=11.7\times10^{22}\mathrm{cm}^{-2}$ in section 1B and then declined back down to  $\mathrm{N}_{\mathrm{H}}\sim4\times10^{22}\mathrm{cm}^{-2}$ through observations two and three. 

Our measurements of column density are in broad agreement with previous observations of IGR J17354$-$3255 (\citet{Tomsick2009atel};\citet{DAi2011}) suggesting that the average absorption column density of this source in quiescence is $\sim4-8\times10^{22}\mathrm{cm}^{-2}$. The expected absorption in the direction of the source is considerably less ($\mathrm{N}_{\mathrm{H}}=1.50\times10^{22}\mathrm{cm}^{-2}$ \citep{Dickey1990}) suggesting that a fraction of the absorbing medium is intrinsic to the binary system. While previous observations of the source have already revealed a high absorbing column density of $(7.5-10)\times10^{22}\mathrm{cm}^{-2}$ (\citet{Tomsick2009atel}, \citet{DAi2011}), the absorption detected between the flares in observation one is still about a factor of two higher than any previously measured in this source. Absorbing column densities of this level have previously been seen in other SFXTs such as IGR J18410$-$0535 \citep{Bozzo2011a} and SAX J1818.6$-$1703 \citep{Boon2016}. The former measured an increase in $\mathrm{N}_{\mathrm{H}}$ from $3-20\times10^{22}\mathrm{cm}^{-2}$ with a further increase to a value of $\sim50\times10^{22}\mathrm{cm}^{-2}$ towards the end of the observation. Their analysis provided evidence that the flare was produced due to the ingestion of a clump of matter. The single incidence of a high absorption column density between two flares seen in observation 1 is most likely caused by changes in the local environment around the compact object rather than any change in the underlying continuum emission mechanism. 

The spectral parameters measured during observations 2 and 3 show minimal variation with intrinsic absorption column density values between $\mathrm{N}_{\mathrm{H}}=8-12\times10^{22}\mathrm{cm}^{-2}$ with the powerlaw model and $\mathrm{N}_{\mathrm{H}}=4-7\times10^{22}\mathrm{cm}^{-2}$ with the blackbody model. Photon index values lie between $\Gamma=\sim1.0-1.5$ and blackbody temperatures are typically of $\sim2$keV. Whilst the blackbody temperatures seem high on first inspection they are consistent with those associated with the surface of an accreting neutron star (\citet{Bartlett2013}; \citet{Lotti2016}).  The same {\scshape{XSPEC}} model used in observation 1 could be utilised to describe the X-ray emission from IGR J17354$-$3255 during all observations, and hence the spectral evolution is most simply explained by a relatively constant underlying continuum shape with changes occurring due to varying absorption.

Observation three exhibits considerable fast flaring, and whilst the absorption column density is near constant, the rapidly varying photon index is likely due to averaging spectra across the rapid changes in flux. A time-resolved spectral analysis was utilised to look into this phenomenon in more detail. The absorption column density was mostly constant around $\sim8\times10^{22}\mathrm{cm}^{-2}$ but even at higher timing resolution, the photon index still varied between 0.9 and 1.6.
Whilst the difference is not great it could indicate changes to the underlying emission mechanism or (more likely considering the timescales involved) could also be explained by the local changes in $\mathrm{N}_{\mathrm{H}}$. Due to the statistical properties of the spectra, however, we are unable to investigate more complex models to explain the behaviour in observation three.

\subsection{Correlated flux - spectral changes?} 

While the power law index was seen to vary significantly through the observations, we were unable to establish any correlations of these changes in photon index with other fit parameters, or with the source brightness. In particular, the  association of a hard spectrum and low flux seen in observation 1B was not seen to hold across the remainder of the dataset, and must be considered a one-off event based on the current dataset.

In a wider context of SFXT behaviour, observation section 1B may be an example of 'pre-flare-hardening' - a sudden increase in the absorption column density together with a decrease in flux associated with subsequent flares. This feature may be seen due to a dense clump of material travelling towards the neutron star during this time. Whilst the $\mathrm{N}_{\mathrm{H}}$ had a significant rise in section 1B, the photon index only showed minimal variability. This phenomenon could be a sign of absorption of a constant intrinsic flux by an optically thin material. Similar behaviour has been observed in SFXTs like IGR J18410$-$0535 \citep{Bozzo2011a} and IGR J16418$-$4532 \citep{Drave2013}, where the resulting conclusion was that an optically thin clump of stellar wind material obscured the intrinsic emission from the neutron star.  The clumpy wind model is often used to interpret the X-ray variability of SFXTs (\citet{In'tZand2005}, \citet{Walter2007}, \citet{Ducci2009b}). The model predicts that fast X-ray flaring can be produced when the neutron star sporadically captures and accretes a clump of stellar wind material. Clumps that pass in front of the X-ray source may cause temporary dimming or obscuration. Therefore we conclude that the absorption features present in Observation 1 may represent structures in the wind i.e. a clump being accreted causing a flare in section 1A, a further optically thin clump of stellar wind material obscuring the intrinsic emission from the neutron star in section 1B and then the latter being captured and accreted in section 1C.

A time-resolved spectral analysis performed at 1000s resolution on observation 3 using a simple powerlaw model did not show any significant correlation between the continuum flux and photon index. This result agrees with the findings of \cite{Bozzo2017} whose analysis of the hardness ratio vs. the flux shows no significant variation. There is therefore no evidence that IGR J17354$-$3255 follows the 'harder-when-brighter' correlation seen in some other sources.

\subsection{The broad-band spectrum}

Extrapolating the mean \XMM flux to an 18-60 keV flux enabled direct comparison with the upper flux limits obtained with the \INTEGRAL analysis. An average flux of $2.6\times10^{-11}$\fluxunits in the 1$-$12 keV band is equivalent $6.65\times10^{-11}$\fluxunits in the 18$-$60 keV band (IBIS/ISGRI) and $1.80\times10^{-11}$\fluxunits in the 3$-$10 keV band (JEM-X). The JEM-X upper limit on the persistent flux was found to be $1.35\times10^{-11}$\fluxunits, similar to the extrapolated flux. Conversely, the extrapolated 18$-$60 keV flux is far in excess of the IBIS/ISGRI upper limit on the persistent flux of $1.00\times10^{-11}$\fluxunits, providing strong evidence that the simple powerlaw spectrum does not extend above 10 keV and consequently a cutoff powerlaw would be needed to explain the IBIS/ISGRI upper limit. It was not possible to constrain a cutoff powerlaw model using the \XMM spectra alone, suggesting that the cutoff likely appears somewhere between 10$-$30 keV. The average \XMM flux for the blackbody model resulted in an 18$-$60 keV flux of $1.3\times10^{-12}$\fluxunits, which is consistent with the IBIS/ISGRI upper limit. Consequently, a single powerlaw can be ruled out as the optimal model to fit to the available \XMM data.

The non-detection of outburst activity by \INTEGRAL is consistent with the particularly low outburst recurrence of this source. \cite{Sguera2011} states that the recurrence rate for emission detectable by INTEGRAL at periastron is only $\sim25\%$.

\subsection{Presence of an iron line?}

\begin{figure}
\minipage{0.45\textwidth}
	\centering
	\includegraphics[width=\columnwidth]{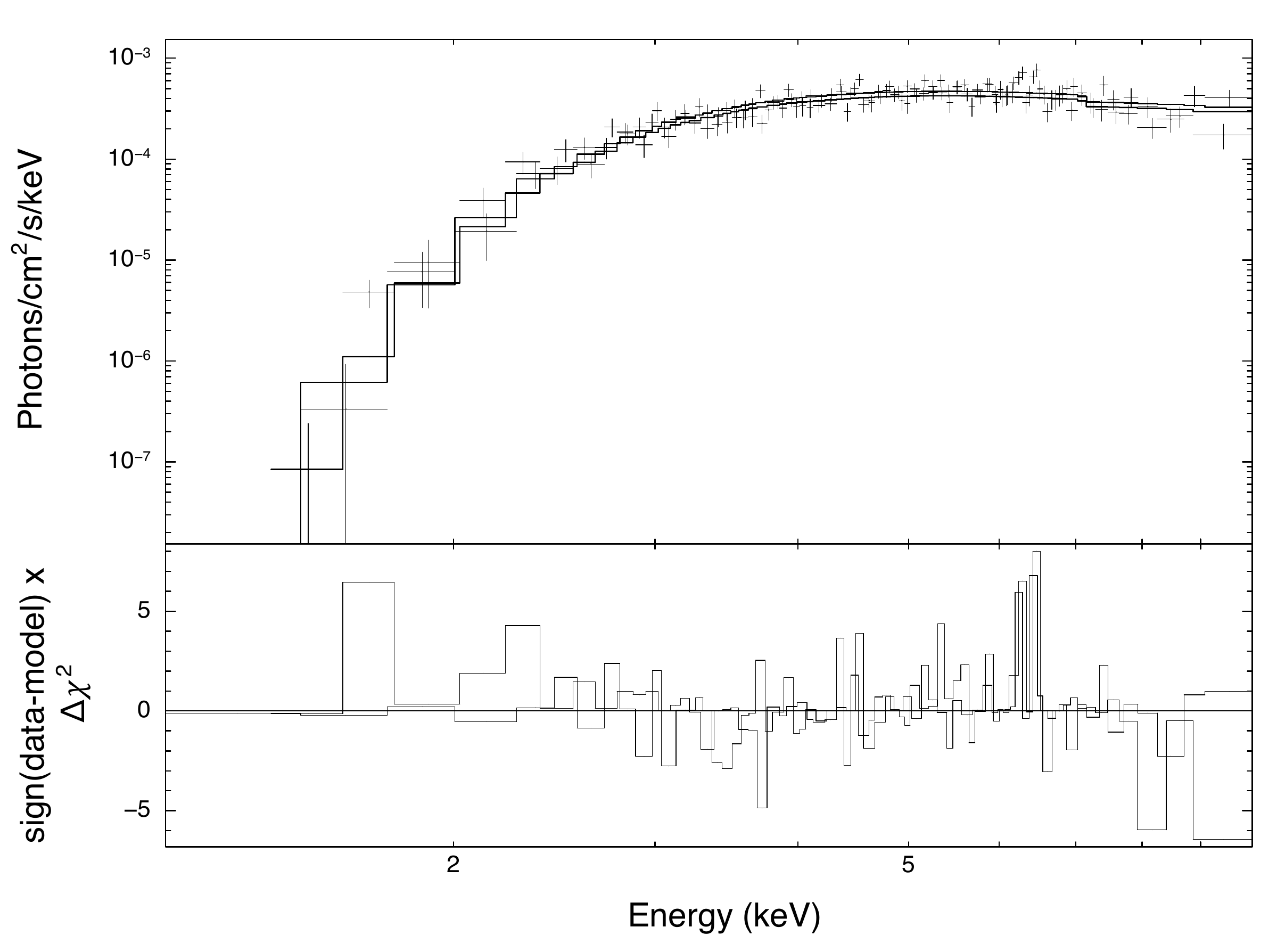}
	
\endminipage\hfill\newline
\minipage{0.45\textwidth}
	\centering
	\includegraphics[width=\columnwidth]{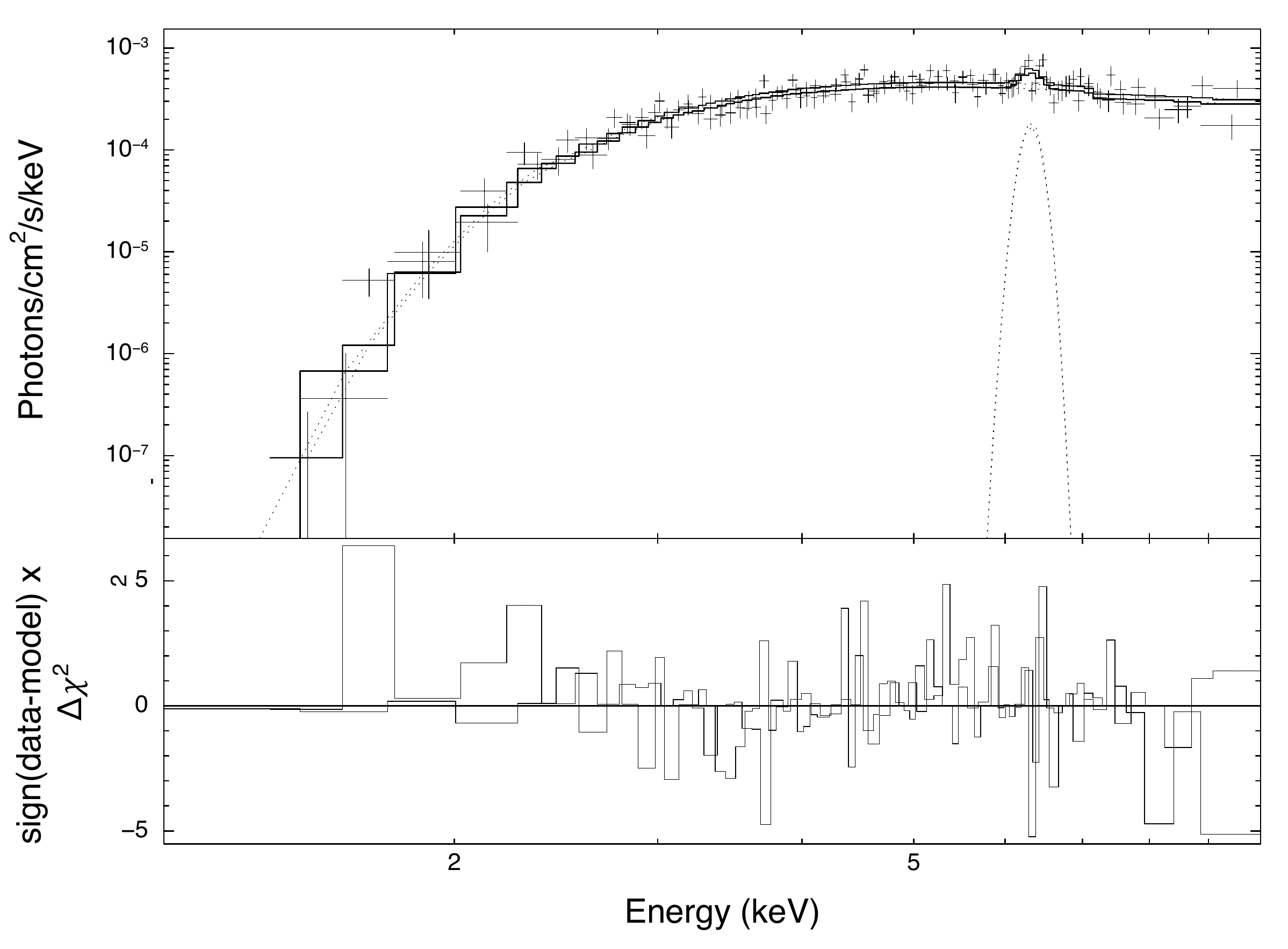}
	
\endminipage
\caption{\label{fig:spectra} Top panel: Blackbody fit with residuals to EPIC-MOS1 and EPIC-MOS2 spectra from observation 1b. Bottom panel: Blackbody fit with an additional Gaussian component from the same observation.}
\end{figure}

In some spectra, residuals in the fit around 6.5 keV hinted at a line feature. Any underlying iron line is likely caused by X-ray irradiation of cold iron in the wind of the supergiant \citep{Kallman2004}. Adding a Gaussian component to account for an iron emission feature improved the fits marginally (but within the errors on the fit statistic) in many cases, and somewhat reduced the residuals in the fit (Figure \ref{fig:spectra}). 
In most cases, we were able to obtain constrained errors on the equivalent widths using the XSPEC {\sc EQWIDTH} tool. In these instances, the line is (as expected from inspection) of marginal significance at best. We note here that the F-test statistic is not suitable to test for the presence of a line, and we can only estimate the significance of the line detection from the errors on the equivalent width, when it is constrained.

%%The centroid energy of the iron line together with its EW and normalisation provides information about the ionisation state of the stellar wind material and the amount of material present around the neutron star. Considering the differences seen in these three observations it leads to the conclusion that the stellar winds in SFXTs can change significantly during the neutron star orbits (e. g. \citet{Walter2007}).

\subsection{Comparison with previous analysis}

The results of our work generally agree with the findings of \cite{Bozzo2017}. They fit an absorbed powerlaw model and achieved reasonable results. However, when also considering the \INTEGRAL upper limit, a blackbody model may be a better fit. Moreover, they found that the addition of an iron line further improved the fits, but we are unable to confirm that in detail due to the statistical quality of the time-resolved spectra we used.

The interpretation that the decrease in flux together with an increase in the absorption column density between the two flares in observation one could be due to a clump of matter obscuring the source further agrees with their results. It is generally considered that simple ingestion of clumps of matter cannot be the sole reason for the high X-ray variability in SFXTs and there must be mechanisms present that inhibit the accretion of the stellar wind material, \cite{Bozzo2017} concluded that IGR J17354$-$3255 seems to have no issues overcoming them. We also see no evidence of spectral changes beyond variable absorption and accretion rate.

\section*{Acknowledgments}
We would like to thank Diego Altamirano, Angela Bazzano and Lara Sidoli for their helpful discussions on the analysis techniques used. This work is based on data from observations with \XMM. \XMM is an ESA science mission with instruments and contributions directly funded by ESA Member States and the USA (NASA). M. E. Goossens is supported by a Mayflower Scholarship from the University of Southampton. A. B. Hill acknowledges that this research was supported by a Marie Curie International Outgoing Fellowship within the 7th European Community Framework Programme (FP7/2007$-$2013) under grant no. 275861. S. P. Drave acknowledges support from the UK Science And Technology Facilities Council, STFC. This research has made use of the SIMBAD database, operated at CDS, Strasbourg, France. This research made use of APLpy, an open-source plotting package for Python \citep{Robitaille2012}.

\bibliographystyle{mnras}
\bibliography{library} % if your bibtex file is called example.bib

\begin{thebibliography}{}
\makeatletter
\relax
\def\mn@urlcharsother{\let\do\@makeother \do\$\do\&\do\#\do\^\do\_\do\%\do\~}
\def\mn@doi{\begingroup\mn@urlcharsother \@ifnextchar [ {\mn@doi@}
  {\mn@doi@[]}}
\def\mn@doi@[#1]#2{\def\@tempa{#1}\ifx\@tempa\@empty \href
  {http://dx.doi.org/#2} {doi:#2}\else \href {http://dx.doi.org/#2} {#1}\fi
  \endgroup}
\def\mn@eprint#1#2{\mn@eprint@#1:#2::\@nil}
\def\mn@eprint@arXiv#1{\href {http://arxiv.org/abs/#1} {{\tt arXiv:#1}}}
\def\mn@eprint@dblp#1{\href {http://dblp.uni-trier.de/rec/bibtex/#1.xml}
  {dblp:#1}}
\def\mn@eprint@#1:#2:#3:#4\@nil{\def\@tempa {#1}\def\@tempb {#2}\def\@tempc
  {#3}\ifx \@tempc \@empty \let \@tempc \@tempb \let \@tempb \@tempa \fi \ifx
  \@tempb \@empty \def\@tempb {arXiv}\fi \@ifundefined
  {mn@eprint@\@tempb}{\@tempb:\@tempc}{\expandafter \expandafter \csname
  mn@eprint@\@tempb\endcsname \expandafter{\@tempc}}}

\bibitem[\protect\citeauthoryear{Bamba, Yokogawa, Ueno, Koyama  \&
  Yamauchi}{Bamba et~al.}{2001}]{Bamba2001}
Bamba A.,  Yokogawa J.,  Ueno M.,  Koyama K.,   Yamauchi S.,  2001, \mn@doi
  [Publications of the Astronomical Society of Japan] {10.1093/pasj/53.6.1179},
  53, 1179

\bibitem[\protect\citeauthoryear{Barthelmy et~al.,}{Barthelmy
  et~al.}{2005}]{Barthelmy2005}
Barthelmy S.~D.,  et~al., 2005, \mn@doi [Space Science Reviews]
  {10.1007/s11214-005-5096-3}, 120, 143

\bibitem[\protect\citeauthoryear{Bartlett, Coe  \& Ho}{Bartlett
  et~al.}{2013}]{Bartlett2013}
Bartlett E.~S.,  Coe M.~J.,   Ho W. C.~G.,  2013, \mn@doi [Monthly Notices of
  the Royal Astronomical Society] {10.1093/mnras/stt1711}, 436, 2054

\bibitem[\protect\citeauthoryear{Baumgartner, Tueller, Markwardt  \&
  Skinner}{Baumgartner et~al.}{2010}]{Baumgartner2010}
Baumgartner W.~H.,  Tueller J.,  Markwardt C.,   Skinner G.,  2010, Bulletin of
  the American Astronomical Society, 42, 675

\bibitem[\protect\citeauthoryear{Bird et~al.,}{Bird et~al.}{2016}]{Bird2016}
Bird A.~J.,  et~al., 2016, \mn@doi [The Astrophysical Journal Supplement
  Series] {10.3847/0067-0049/223/1/15}, 223, 15

\bibitem[\protect\citeauthoryear{Boon et~al.,}{Boon et~al.}{2016}]{Boon2016}
Boon C.~M.,  et~al., 2016, Monthly Notices of the Royal Astronomical Society,
  456, 4111

\bibitem[\protect\citeauthoryear{Bozzo, Falanga  \& Stella}{Bozzo
  et~al.}{2008}]{Bozzo2008b}
Bozzo E.,  Falanga M.,   Stella L.,  2008, in proceedings of "7th INTEGRAL
  Workshop", Sep., PoS(INTEGRAL2008)101

\bibitem[\protect\citeauthoryear{Bozzo et~al.,}{Bozzo
  et~al.}{2011}]{Bozzo2011a}
Bozzo E.,  et~al., 2011, \mn@doi [Astronomy {\&} Astrophysics]
  {10.1051/0004-6361/201116726}, 531, A130

\bibitem[\protect\citeauthoryear{Bozzo, Pavan, Ferrigno, Falanga, Campana,
  Paltani, Stella  \& Walter}{Bozzo et~al.}{2012}]{Bozzo2012b}
Bozzo E.,  Pavan L.,  Ferrigno C.,  Falanga M.,  Campana S.,  Paltani S.,
  Stella L.,   Walter R.,  2012, \mn@doi [Astronomy {\&} Astrophysics]
  {10.1051/0004-6361/201218900}, 544, A118

\bibitem[\protect\citeauthoryear{Bozzo, Bernardini, Ferrigno, Falanga, Romano
  \& Oskinova}{Bozzo et~al.}{2017}]{Bozzo2017}
Bozzo E.,  Bernardini F.,  Ferrigno C.,  Falanga M.,  Romano P.,   Oskinova L.,
   2017, Astronomy {\&} Astrophysics, 608, A128

\bibitem[\protect\citeauthoryear{{Buccheri} et~al.,}{{Buccheri}
  et~al.}{1983}]{z-squared}
{Buccheri} R.,  et~al., 1983, \aap, \href
  {http://adsabs.harvard.edu/abs/1983A%26A...128..245B} {128, 245}

\bibitem[\protect\citeauthoryear{Bulgarelli, Gianotti, Trifoglio  \&
  Striani}{Bulgarelli et~al.}{2009}]{Bulgarelli2009}
Bulgarelli A.,  Gianotti F.,  Trifoglio M.,   Striani E.,  2009, {ATel
  {\#}2017: The candidate gamma-ray transient AGL J1734-3310 in the Galactic
  plane.}

\bibitem[\protect\citeauthoryear{Clark et~al.,}{Clark
  et~al.}{2010}]{Clark2010a}
Clark D.~J.,  et~al., 2010, \mn@doi [Monthly Notices of the Royal Astronomical
  Society: Letters] {10.1111/j.1745-3933.2010.00885.x}, 406, L75

\bibitem[\protect\citeauthoryear{Coleiro, Chaty, Heras, Rahoui  \&
  Tomsick}{Coleiro et~al.}{2013}]{Coleiro2013a}
Coleiro A.,  Chaty S.,  Heras J. A.~Z.,  Rahoui F.,   Tomsick J.~A.,  2013,
  Astronomy {\&} Astrophysics, 560, A108

\bibitem[\protect\citeauthoryear{Cusumano et~al.,}{Cusumano
  et~al.}{2010}]{Cusumano2010}
Cusumano G.,  et~al., 2010, \mn@doi [Astronomy and Astrophysics]
  {10.1051/0004-6361/201015249}, 524, A64

\bibitem[\protect\citeauthoryear{D'A{\`{i}}, {La Parola}, Cusumano, Segreto,
  Romano, Vercellone  \& Robba}{D'A{\`{i}} et~al.}{2011}]{DAi2011}
D'A{\`{i}} A.,  {La Parola} V.,  Cusumano G.,  Segreto A.,  Romano P.,
  Vercellone S.,   Robba N.~R.,  2011, \mn@doi [Astronomy {\&} Astrophysics]
  {10.1051/0004-6361/201016401}, 529, A30

\bibitem[\protect\citeauthoryear{Dickey \& Lockman}{Dickey \&
  Lockman}{1990}]{Dickey1990}
Dickey J.~M.,  Lockman F.~J.,  1990, \mn@doi [Annual Review of Astronomy and
  Astrophysics] {10.1146/annurev.aa.28.090190.001243}, 28, 215

\bibitem[\protect\citeauthoryear{Drave, Bird, Sidoli, Sguera, McBride, Hill,
  Bazzano  \& Goossens}{Drave et~al.}{2013}]{Drave2013}
Drave S.~P.,  Bird A.~J.,  Sidoli L.,  Sguera V.,  McBride V.~A.,  Hill A.~B.,
  Bazzano A.,   Goossens M.~E.,  2013, Monthly Notices of the Royal
  Astronomical Society, 433, 528

\bibitem[\protect\citeauthoryear{Ducci, Sidoli, Mereghetti, Paizis  \&
  Romano}{Ducci et~al.}{2009}]{Ducci2009b}
Ducci L.,  Sidoli L.,  Mereghetti S.,  Paizis A.,   Romano P.,  2009, \mn@doi
  [Monthly Notices of the Royal Astronomical Society]
  {10.1111/j.1365-2966.2009.15265.x}, 398, 2152

\bibitem[\protect\citeauthoryear{Ducci, Romano, Esposito, Bozzo, Krimm,
  Vercellone, Mangano  \& Kennea}{Ducci et~al.}{2013}]{Ducci2013}
Ducci L.,  Romano P.,  Esposito P.,  Bozzo E.,  Krimm H.~A.,  Vercellone S.,
  Mangano V.,   Kennea J.~A.,  2013, \mn@doi [Astronomy {\&} Astrophysics]
  {10.1051/0004-6361/201321635}, 556, A72

\bibitem[\protect\citeauthoryear{Gabriel et~al.,}{Gabriel
  et~al.}{2004}]{Gabriel2004}
Gabriel C.,  et~al., 2004, Astronomical Data Analysis Software and Systems
  (ADASS) XIII, 314, 759

\bibitem[\protect\citeauthoryear{Hill}{Hill}{2016}]{Hill2016}
Hill A.,  2016, \mn@doi [bayesBlocks: First alpha release]
  {10.5281/ZENODO.57750}, Zenodo, DOI:10.5281/ZENODO.57750

\bibitem[\protect\citeauthoryear{Jethwa, Saxton, Guainazzi, Rodriguez-Pascual
  \& Stuhlinger}{Jethwa et~al.}{2015}]{Jethwa2015}
Jethwa P.,  Saxton R.,  Guainazzi M.,  Rodriguez-Pascual P.,   Stuhlinger M.,
  2015, \mn@doi [A{\&}A] {10.1051/0004-6361/201425579}, 581

\bibitem[\protect\citeauthoryear{Kallman, Palmeri, Bautista, Mendoza  \&
  Krolik}{Kallman et~al.}{2004}]{Kallman2004}
Kallman T.~R.,  Palmeri P.,  Bautista M.~A.,  Mendoza C.,   Krolik J.~H.,
  2004, \mn@doi [The Astrophysical Journal Supplement Series] {10.1086/424039},
  155, 675

\bibitem[\protect\citeauthoryear{Kuulkers et~al.,}{Kuulkers
  et~al.}{2006}]{Kuulkers}
Kuulkers E.,  et~al., 2006, {ATel {\#}874: New INTEGRAL source, IGR
  J17354-3255, and continuation of the INTEGRAL Galactic Bulge monitoring
  program}

\bibitem[\protect\citeauthoryear{Kuulkers et~al.,}{Kuulkers
  et~al.}{2007}]{Kuulkers2007}
Kuulkers E.,  et~al., 2007, \mn@doi [Astronomy and Astrophysics]
  {10.1051/0004-6361:20066651}, 466, 595

\bibitem[\protect\citeauthoryear{Leahy, Darbro, Elsner, Weisskopf, Kahn,
  Sutherland  \& Grindlay}{Leahy et~al.}{1983}]{Leahy1983}
Leahy D.~A.,  Darbro W.,  Elsner R.~F.,  Weisskopf M.~C.,  Kahn S.,  Sutherland
  P.~G.,   Grindlay J.~E.,  1983, \mn@doi [The Astrophysical Journal]
  {10.1086/160766}, 266, 160

\bibitem[\protect\citeauthoryear{Lomb}{Lomb}{1976}]{Lomb1976}
Lomb N.~R.,  1976, \mn@doi [Astrophysics and Space Science]
  {10.1007/BF00648343}, 39, 447

\bibitem[\protect\citeauthoryear{Lotti et~al.,}{Lotti et~al.}{2016}]{Lotti2016}
Lotti S.,  et~al., 2016, \mn@doi [The Astrophysical Journal]
  {10.3847/0004-637X/822/1/57}, 822, 57

\bibitem[\protect\citeauthoryear{Lund et~al.,}{Lund et~al.}{2003}]{Lund2003}
Lund N.,  et~al., 2003, \mn@doi [Astronomy and Astrophysics]
  {10.1051/0004-6361:20031358}, 411, L231

\bibitem[\protect\citeauthoryear{Negueruela, Smith, Reig, Chaty  \&
  Torrejon}{Negueruela et~al.}{2006}]{Negueruela2005}
Negueruela I.,  Smith D.~M.,  Reig P.,  Chaty S.,   Torrejon J.~M.,  2006, The
  X-ray Universe 2005, 604, 165

\bibitem[\protect\citeauthoryear{Press \& Rybicki}{Press \&
  Rybicki}{1989}]{Press1989}
Press W.~H.,  Rybicki G.~B.,  1989, \mn@doi [The Astrophysical Journal]
  {10.1086/167197}, 338, 277

\bibitem[\protect\citeauthoryear{Read, Guainazzi  \& Sembay}{Read
  et~al.}{2014}]{Read2014a}
Read A.~M.,  Guainazzi M.,   Sembay S.,  2014, \mn@doi [Astronomy {\&}
  Astrophysics] {10.1051/0004-6361/201423422}, 564, A75

\bibitem[\protect\citeauthoryear{Robitaille \& Bressert}{Robitaille \&
  Bressert}{2012}]{Robitaille2012}
Robitaille T.,  Bressert E.,  2012, Astrophysics Source Code Library

\bibitem[\protect\citeauthoryear{Romano et~al.,}{Romano
  et~al.}{2015}]{Romano2015}
Romano P.,  et~al., 2015, \mn@doi [Astronomy {\&} Astrophysics]
  {10.1051/0004-6361/201525749}, 576, L4

\bibitem[\protect\citeauthoryear{Scargle}{Scargle}{1982}]{Scargle1982}
Scargle J.~D.,  1982, \mn@doi [The Astrophysical Journal] {10.1086/160554},
  263, 835

\bibitem[\protect\citeauthoryear{Scargle, Norris, Jackson  \& Chiang}{Scargle
  et~al.}{2013}]{Scargle2013}
Scargle J.~D.,  Norris J.~P.,  Jackson B.,   Chiang J.,  2013, \mn@doi [The
  Astrophysical Journal] {10.1088/0004-637X/764/2/167}, 764, 167

\bibitem[\protect\citeauthoryear{Sguera}{Sguera}{2013}]{Sguera2013}
Sguera V.,  2013, Nuclar Physics B -Proceedings Supplements, 239, 76

\bibitem[\protect\citeauthoryear{Sguera et~al.,}{Sguera
  et~al.}{2005}]{Sguera2005}
Sguera V.,  et~al., 2005, \mn@doi [Astronomy and Astrophysics]
  {10.1051/0004-6361:20053103}, 444, 221

\bibitem[\protect\citeauthoryear{Sguera et~al.,}{Sguera
  et~al.}{2006}]{Sguera2006}
Sguera V.,  et~al., 2006, \mn@doi [The Astrophysical Journal] {10.1086/504827},
  646, 452

\bibitem[\protect\citeauthoryear{Sguera et~al.,}{Sguera
  et~al.}{2007}]{Sguera2007}
Sguera V.,  et~al., 2007, \mn@doi [Astronomy {\&} Astrophysics]
  {10.1051/0004-6361}, 462, 695

\bibitem[\protect\citeauthoryear{Sguera, Drave, Bird, Bazzano, Landi  \&
  Ubertini}{Sguera et~al.}{2011}]{Sguera2011}
Sguera V.,  Drave S.~P.,  Bird A.~J.,  Bazzano A.,  Landi R.,   Ubertini P.,
  2011, \mn@doi [Monthly Notices of the Royal Astronomical Society]
  {10.1111/j.1365-2966.2011.19298.x}, 417, 573

\bibitem[\protect\citeauthoryear{Shakura, Postnov, Sidoli  \& Paizis}{Shakura
  et~al.}{2014}]{Shakura2014}
Shakura N.,  Postnov K.,  Sidoli L.,   Paizis A.,  2014, \mn@doi [Monthly
  Notices of the Royal Astronomical Society] {10.1093/mnras/stu1027}, 442, 2325

\bibitem[\protect\citeauthoryear{Sidoli, Mereghetti, Sguera  \&
  Pizzolato}{Sidoli et~al.}{2012}]{Sidoli2012a}
Sidoli L.,  Mereghetti S.,  Sguera V.,   Pizzolato F.,  2012, \mn@doi [Monthly
  Notices of the Royal Astronomical Society]
  {10.1111/j.1365-2966.2011.20063.x}, 420, 554

\bibitem[\protect\citeauthoryear{Str{\"{u}}der et~al.,}{Str{\"{u}}der
  et~al.}{2001}]{Strder2001a}
Str{\"{u}}der L.,  et~al., 2001, \mn@doi [Astronomy and Astrophysics]
  {10.1051/0004-6361:20000066}, 365, L18

\bibitem[\protect\citeauthoryear{Tomsick}{Tomsick}{2009}]{Tomsick2009atel}
Tomsick J.~A.,  2009, {ATel {\#}2022: A Chandra X-ray localization, spectrum,
  and IR identification for IGR J17354-3255}

\bibitem[\protect\citeauthoryear{Tomsick, Chaty, Rodriguez, Walter  \&
  Kaaret}{Tomsick et~al.}{2009a}]{Tomsick2009c}
Tomsick J.~A.,  Chaty S.,  Rodriguez J.,  Walter R.,   Kaaret P.,  2009a,
  \mn@doi [The Astrophysical Journal, Volume 701, Issue 1, pp. 811-823 (2009).]
  {10.1088/0004-637X/701/1/811}, 701, 811

\bibitem[\protect\citeauthoryear{Tomsick, Chaty, Rodriguez, Walter  \&
  Kaaret}{Tomsick et~al.}{2009b}]{Tomsick2009}
Tomsick J.~A.,  Chaty S.,  Rodriguez J.,  Walter R.,   Kaaret P.,  2009b,
  \mn@doi [The Astrophysical Journal] {10.1088/0004-637X/701/1/811}, 701, 811

\bibitem[\protect\citeauthoryear{Turner et~al.,}{Turner
  et~al.}{2001}]{Turner2001}
Turner M. J.~L.,  et~al., 2001, \mn@doi [Astronomy and Astrophysics]
  {10.1051/0004-6361:20000087}, 365, L27

\bibitem[\protect\citeauthoryear{Ubertini et~al.,}{Ubertini
  et~al.}{2003}]{Ubertini2003}
Ubertini P.,  et~al., 2003, \mn@doi [Astronomy and Astrophysics]
  {10.1051/0004-6361:20031224}, 411, L131

\bibitem[\protect\citeauthoryear{Vedrenne et~al.,}{Vedrenne
  et~al.}{2003}]{Vedrenne2003}
Vedrenne G.,  et~al., 2003, \mn@doi [Astronomy and Astrophysics]
  {10.1051/0004-6361:20031482}, 411, L63

\bibitem[\protect\citeauthoryear{Walter \& {Zurita Heras}}{Walter \& {Zurita
  Heras}}{2007}]{Walter2007}
Walter R.,  {Zurita Heras} J.,  2007, \mn@doi [Astronomy and Astrophysics]
  {10.1051/0004-6361:20078353}, 476, 335

\bibitem[\protect\citeauthoryear{Wilms, Allen  \& McCray}{Wilms
  et~al.}{2000}]{Wilms2000}
Wilms J.,  Allen A.,   McCray R.,  2000, \mn@doi [The Astrophysical Journal]
  {10.1086/317016}, 542, 914

\bibitem[\protect\citeauthoryear{den Herder et~al.,}{den Herder
  et~al.}{2001}]{DenHerder2001}
den Herder J.~W.,  et~al., 2001, \mn@doi [Astronomy and Astrophysics]
  {10.1051/0004-6361:20000058}, 365, L7

\bibitem[\protect\citeauthoryear{{in 't Zand}}{{in 't
  Zand}}{2005}]{In'tZand2005}
{in 't Zand} J. J.~M.,  2005, \mn@doi [Astronomy and Astrophysics]
  {10.1051/0004-6361:200500162}, 441, L1

\makeatother
\end{thebibliography}

%\begin{thebibliography}{99}
%\end{thebibliography}

%%%%%%%%%%%%%%%%%%%%%%%%%%%%%%%%%%%%%%%%%%%%%%%%%%

%%%%%%%%%%%%%%%%% APPENDICES %%%%%%%%%%%%%%%%%%%%%

\appendix
\section{Additional time-resolved spectral analysis using a {\scshape tbabs*powerlaw model}}

\begin{table*}
\begin{minipage}{170mm}
\centering
\caption{Time-resolved spectral analysis with a resolution of 1000s of observation three using the tbabs*powerlaw model. The model parameters as well as their lower and upper limits are displayed. }
\label{table:six}
\resizebox{\columnwidth}{!}{
\begin{tabular}{llllllllllll}
\hline
\hline
Number & $\chi_{\mathrm{red}}^{2}$ & dof & $\mathrm{N}_{\mathrm{H}}$    & $\mathrm{N}_{\mathrm{H}}$LL & $\mathrm{N}_{\mathrm{H}}$UL  & Gamma & GammaLL & GammaUL & Flux & FluxLL   & FluxUL   \\
&&&$[\times10^{22}\mathrm{cm}^{-2}]$&$[\times10^{22}\mathrm{cm}^{-2}]$&$[\times10^{22}\mathrm{cm}^{-2}]$&&&&[erg $\mathrm{cm}^{-2}\mathrm{s}^{-1}$]&[erg $\mathrm{cm}^{-2}\mathrm{s}^{-1}$]&[erg $\mathrm{cm}^{-2}\mathrm{s}^{-1}$]\\
\hline
1      & 0.97   & 72  & 8.18  & 6.76 & 9.81  & 1.44  & 1.16    & 1.75    & 8.05E-12   & 6.51E-12 & 8.43E-12 \\
2      & 1.07   & 184 & 7.53  & 6.73 & 8.41  & 1.47  & 1.32    & 1.63    & 1.37E-11   & 1.28E-11 & 1.40E-11 \\
3      & 1.06   & 218 & 8.13  & 7.43 & 8.90  & 1.54  & 1.42    & 1.67    & 1.70E-11   & 1.63E-11 & 1.74E-11 \\
4      & 1.00   & 172 & 9.46  & 8.49 & 10.54 & 1.57  & 1.41    & 1.74    & 1.12E-11   & 1.05E-11 & 1.16E-11 \\
5      & 1.47   & 182 & 8.73  & 7.88 & 9.68  & 1.28  & 1.14    & 1.43    & 1.28E-11   & 1.21E-11 & 1.31E-11 \\
6      & 1.38   & 189 & 8.26  & 7.43 & 9.17  & 1.26  & 1.12    & 1.40    & 1.33E-11   & 1.26E-11 & 1.36E-11 \\
7      & 1.13   & 126 & 7.81  & 6.81 & 8.93  & 1.40  & 1.22    & 1.60    & 7.18E-12   & 6.56E-12 & 7.45E-12 \\
8      & 0.99   & 138 & 7.59  & 6.65 & 8.65  & 1.34  & 1.16    & 1.52    & 8.25E-12   & 7.62E-12 & 8.52E-12 \\
9      & 1.06   & 110 & 7.64  & 6.38 & 9.13  & 1.17  & 0.97    & 1.39    & 6.68E-12   & 6.10E-12 & 6.97E-12 \\
10     & 0.97   & 170 & 7.72  & 6.85 & 8.68  & 1.15  & 1.01    & 1.30    & 1.17E-11   & 1.10E-11 & 1.21E-11 \\
11     & 1.02   & 222 & 8.01  & 7.30 & 8.78  & 1.22  & 1.10    & 1.34    & 1.78E-11   & 1.70E-11 & 1.82E-11 \\
12     & 1.05   & 241 & 8.52  & 7.87 & 9.23  & 1.33  & 1.22    & 1.44    & 2.33E-11   & 2.24E-11 & 2.38E-11 \\
13     & 1.13   & 255 & 8.47  & 7.80 & 9.19  & 1.15  & 1.05    & 1.25    & 2.73E-11   & 2.64E-11 & 2.79E-11 \\
14     & 1.17   & 274 & 9.12  & 8.41 & 9.88  & 0.89  & 0.79    & 0.98    & 3.46E-11   & 3.35E-11 & 3.53E-11 \\
15     & 1.21   & 304 & 8.56  & 8.00 & 9.17  & 0.93  & 0.85    & 1.01    & 4.28E-11   & 4.17E-11 & 4.37E-11 \\
16     & 1.13   & 286 & 9.07  & 8.47 & 9.72  & 1.12  & 1.03    & 1.21    & 3.70E-11   & 3.60E-11 & 3.77E-11 \\
17     & 1.18   & 253 & 9.73  & 8.98 & 10.54 & 1.20  & 1.09    & 1.31    & 2.72E-11   & 2.62E-11 & 2.78E-11 \\
18     & 1.06   & 217 & 9.34  & 8.48 & 10.29 & 1.17  & 1.05    & 1.30    & 1.83E-11   & 1.77E-11 & 1.87E-11 \\
19     & 1.02   & 195 & 10.20 & 9.18 & 11.34 & 1.30  & 1.16    & 1.44    & 1.53E-11   & 1.44E-11 & 1.57E-11 \\
20     & 1.07   & 241 & 9.38  & 8.65 & 10.17 & 1.43  & 1.32    & 1.55    & 2.06E-11   & 1.97E-11 & 2.10E-11 \\
21     & 1.03   & 243 & 8.24  & 7.63 & 8.90  & 1.40  & 1.29    & 1.51    & 2.75E-11   & 2.64E-11 & 2.82E-11\\
\hline
\end{tabular}}
\end{minipage}
\end{table*}


%%%%%%%%%%%%%%%%%%%%%%%%%%%%%%%%%%%%%%%%%%%%%%%%%%

% Don't change these lines
\bsp	% typesetting comment
\label{lastpage}
\end{document}